\documentclass[11pt,a4paper, final]{article}

\usepackage[a4paper, left=22mm, right=22mm]{geometry}
\usepackage[T1]{fontenc}							
\usepackage[utf8]{inputenc}							
\usepackage{amsmath, amsfonts, amssymb, amsthm, empheq}		
\usepackage{graphicx}								
\usepackage{microtype}								
\usepackage{hyperref}								
\usepackage{cleveref}								
\usepackage{booktabs}								
\usepackage{todonotes}
\usepackage{float}
\usepackage{caption}
\usepackage{subcaption}						
\usepackage[notref,notcite]{showkeys}				
\usepackage{siunitx}
\usepackage{array} 
\usepackage{amssymb}
\usepackage{amsmath, amsfonts, amssymb, amsthm}
\usepackage{titling}
\usepackage[
  sortlocale=auto,  
  giveninits=true, 
  bibencoding=auto, 
  isbn=false,  
  url=false, 
  doi=false, 
  autocite=superscript,
  eprint=false,  
  useprefix=false,  
  sorting=none,  
  backref=false, 
  minnames=1, 
  maxcitenames=3, 
  maxbibnames=10, 
  defernumbers = true
]{biblatex}

\DeclareCiteCommand{\citenum}
  {}
  {\printfield{labelnumber}}
  {}
  {}

\usepackage{filecontents}

\usepackage{xr}
\def\electrons{\text{n}}
\def\holes{\text{p}}

\def\d{\mathrm{d}}

\def\JSC{J_\text{SC}}
\def\VOC{V_\text{OC}}

\def\cms{\,cm$\cdot$s$^{-1}$}

\thanksmarkseries{arabic}

\newcommand*\samethanks[1][\value{footnote}]{\footnotemark[#1]} 

\title{How nanotextured interfaces influence the electronics in perovskite solar cells}

\author{
    Dilara Abdel\thanks{Numerical Methods for Innovative Semiconductor Devices, Weierstrass Institute for Applied Analysis and Stochastics (WIAS), Mohrenstr. 39, 10117 Berlin, Germany. E-mail: patricio.farrell@wias-berlin.de}
    \and Jacob Relle\thanks{Optics for Solar Energy, Helmholtz-Zentrum Berlin für Materialien und Energie GmbH, Berlin, Germany.
    E-mail: christiane.becker@helmholtz-berlin.de, klaus.jaeger@helmholtz-berlin.de.}
    $^{,}$\thanks{Computational Nano Optics, Zuse Institute Berlin, Berlin, Germany.}
    \and Thomas Kirchartz\thanks{IMD-3 Photovoltaics, Forschungszentrum Jülich GmbH, J\"ulich, Germany.}
    $^{,}$\thanks{University of Duisburg-Essen, Duisburg, Germany.}
    \and Patrick Jaap\thanks{Numerical Mathematics and Scientific Computing, Weierstrass Institute for Applied Analysis and Stochastics (WIAS), Berlin, Germany.}
    \and Jürgen Fuhrmann\samethanks[6]
    \and Sven Burger\samethanks[3]
    $^{,}$\thanks{JCMwave GmbH, Berlin, Germany.}
    \and Christiane Becker\samethanks[2]
    $^{,}$\thanks{Hochschule für Technik und Wirtschaft Berlin, Berlin, Germany.}
    \and Klaus Jäger\samethanks[2] $^{,}$\samethanks[3]
    \and Patricio Farrell\samethanks[1]
}

\date{\today}

\begin{document}

    \maketitle
    \begin{abstract}
      Perovskite solar cells have reached power conversion efficiencies that rival those of established silicon photovoltaics.
      Nanotextures in perovskite solar cells scatter the incident light, thereby improving optical absorption.
      In addition, experiments show that nanotextures impact electronic performance, although the underlying mechanisms remain unclear.
      This study investigates the underlying theoretical reasons by combining multi-dimensional optical and charge-transport simulations for a single-junction perovskite solar cell.
      Our numerical results reveal that texturing redistributes the electric field, influencing carrier accumulation and recombination dynamics.
      We find that moderate texturing heights ($\leq 300$ nm) always increase the power conversion efficiency, regardless of surface recombination velocities.
      Our study also clarifies why experiments have reported that texturing both increased and reduced open-circuit voltages in perovskite solar cells: this behaviour originates from variations in surface recombination at the untextured electron transport layer.
      In contrast, surface recombination at the textured hole transport layer strongly affects the short-circuit current density, with lower recombination rates keeping it closer to the optical ideal.
      These findings provide new insights into the opto-electronic advantages of texturing and offer guidance for the design of next-generation textured perovskite-based solar cells, light emitting diodes, and photodetectors.
    \end{abstract}

\subsection*{Introduction}

In recent years, perovskite-based solar cells have rapidly advanced photovoltaics by combining high power conversion efficiencies (PCEs) with low-cost and scalable fabrication methods.
Their outstanding opto-electronic properties, such as tunable band gaps or strong absorption, make them highly attractive for next-generation solar energy applications.\autocite{Han2025, bati2023next}

Multi-junction solar cells, comprising multiple sub-cells with different bandgaps, mitigate thermalisation losses and thereby enhance the PCE (see, e.g., Ref. \citenum{Jaeger2025}, Chapter 7).
In tandem solar cells two junctions are combined.
Both all-perovskite and perovskite-silicon tandem devices have surpassed the efficiency limit of traditional single-junction silicon cells.\autocite{wang2024homogenized, liu2024perovskite, liu2025all, wang2025highly, ugur2024enhanced}
These improvements are driven by advances in material composition, interface passivation, and increasingly sophisticated device architectures.

Introducing textured interfaces is a commonly used strategy to further enhance device performance.
In most studies on perovskite solar cells, texturing has been mainly motivated and discussed from an optical perspective.\autocite{paetzold:2015apl,Sahli_2018,Chen_2020,Hou_2020,tockhorn2020improved}
Pioneering studies have demonstrated enhanced light absorption and short-circuit current densities $\JSC$ in perovskite-silicon tandem devices, featuring pyramidal textures on the micrometer and sub-micrometer scale.\autocite{Sahli_2018, Chen_2020,Hou_2020}
However, the texture-related increase of $J_{\text{SC}}$ was often accompanied by a reduced electronic performance in terms of open-circuit voltage $V_{\text{OC}}$.\autocite{paetzold:2015apl,Winarto2025}

Beyond these optically motivated and experimentally verified enhancements in short-circuit current density $\JSC$, several experimental studies have also reported \textit{increases} of the open-circuit voltage $\VOC$ in textured perovskite solar cells.
In single-junction devices, for instance, $\VOC$ gains of up to $+20$\,mV have been measured, which cannot be explained by the weak logarithmic dependence of $\VOC$ on $\JSC$.\autocite{tockhorn2020improved, liu2022efficient}
Similarly, beneficial but unexplained voltage gains between $+15$\,mV and $+45$\,mV have been reported in tandem architectures.\autocite{tockhorn2022nano, Hou_2020, zheng2023balancing}
Proposed mechanisms include improved charge carrier collection due to a widened depletion region\autocite{Hou_2020} and suppressed non-radiative recombination\autocite{er2023loss}.
However, a detailed physical understanding of the enhancement or reduction of $V_{\text{OC}}$ in textured devices is still lacking.
As interfaces are known to play a crucial role on non-radiative recombination and voltage losses in perovskite solar cells,\autocite{Wolff2019}, surface-enlarging textures will have a significant impact on the $\VOC$  of the devices.

Opto-electronic simulations can help to study perovskite solar cells, including vacancy migration and advanced light absorption models. Often, commercial software tools are used,\autocite{Neukom2019, Aeberhard24, messmer2025toward} which typically lack the flexibility to implement customised physical models.
As an alternative, (partially) open-source simulation tools have been developed to study vacancy-assisted charge transport.\autocite{Clarke2022, Calado2022, Koopmans2022simsalabim, tahir2025computational}
However, such one-dimensional approaches are inherently limited in their ability to capture the spatial effects introduced by nanoscale textures.
Recent multi-dimensional studies have begun to simulate textured perovskite architectures, particularly in tandem devices.
Still, these works primarily focus on device optimization, without offering insight into underlying physics.\autocite{er2023loss, hsieh2024analysis, huang2019optimization}

In this work, we present a multi-dimensional simulation framework, applied here in two spatial dimensions, which couples optical finite element simulations of the time-harmonic Maxwell equations via \texttt{JCMsuite}\autocite{Pomplun2007pssb} with electronic finite volume simulations using \texttt{ChargeTransport.jl},\autocite{ChargeTransport,Abdel2021Model} which solve the drift-diffusion equations.
The optical simulations allow us to generate geometry-dependent photogeneration profiles, which are passed to the electronic solver that calculates coupled electronic-ionic charge carrier transport in the textured perovskite solar cell.
This integrated approach allows us to quantify how nanotextures influence carrier dynamics.

The remainder of this paper is structured as follows:
We first present a well-studied planar single-junction perovskite solar cell architecture from the literature,\autocite{LeCorre2022, thiesbrummel2024ion,Diekmann2021} which serves as the baseline for this study.
Then, we describe the extensions introduced to model the textured cells.
Finally, we analyse the influence of texture height on light absorption and key performance indicators, including short-circuit current density, open-circuit voltage, and power conversion efficiency.
To uncover the mechanism behind the enhanced efficiencies, we identify the dominant recombination mechanism, and demonstrate that a redistribution of the electric field governs the electronic response.

\subsection*{Results and discussion}

\subsubsection*{Perovskite solar cell simulation setup}\label{Setup}

The motivation for this study are two seemingly contradictory experimental findings: nanotextured perovskite solar cells have been reported to show both increased and reduced open-circuit voltages for different devices.\autocite{paetzold:2015apl, tockhorn2020improved, liu2022efficient, Winarto2025}

\begin{figure*}[ht!]
  \centering
  \includegraphics[width=\textwidth]{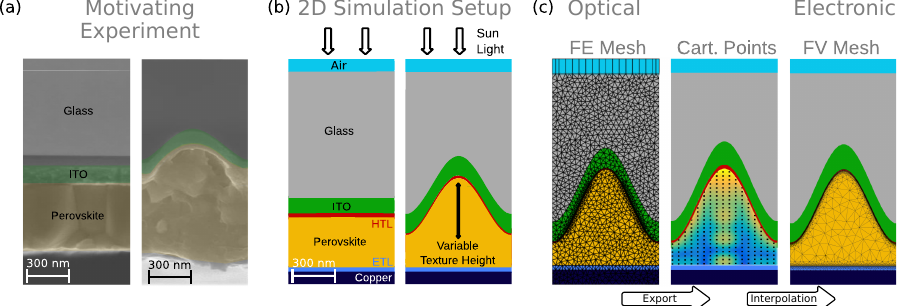}
  \caption{
  Overview of motivating experimental devices, simulated stack and simulation methods used in this study.
  (a) Scanning electron microscope (SEM) cross-section micrograph of the physical layer stacks investigated by Tockhorn \emph{et al.}\ \autocite{tockhorn2020improved} and Sutter,\autocite{Sutter22} showing planar (left) and textured (right) substrates.
  These devices showed improved efficiencies, which inspired the present work.
  The hole transport layer (HTL), electron transport layer (ETL), and the copper back contact layer are not visible.
  (b) The theoretical setup used for the optical and electronic simulations, depicting both planar (left) and textured (right) configurations.
  The material composition varies from the stacks in (a) and is taken from the literature\autocite{Diekmann2021,LeCorre2022,thiesbrummel2024ion} due to available electronic data from these sources.
  (c) Schematic illustration of the three-step coupling procedure between optical and electronic simulations: A finite element (FE) mesh for the optical simulation (left), Cartesian grid points used for exporting the photogeneration rate (middle), and a finite volume (FV) mesh for the electronic simulation (right).
  \label{fig:1}
  }
\end{figure*}

Figure \ref{fig:1}a shows cross-sectional scanning electron microscope (SEM) images of representative planar and sinusoidally textured substrates from previous studies.\autocite{tockhorn2020improved, Sutter22}
Motivated by these experiments, our theoretical investigation is based on a well-characterised planar single-junction perovskite solar cell architecture,\autocite{LeCorre2022, thiesbrummel2024ion,Diekmann2021} for which detailed electronic data were reported.
We extend this planar setup by incorporating two-dimensional (2D) sinusoidal nanotextures.
The layer stack shown in Fig.\ \ref{fig:1}b (left) consists of (from top to bottom) a glass substrate, an indium tin oxide (ITO) front electrode, a hole transport layer (HTL), a perovskite absorber (PVK), an electron transport layer (ETL), and a copper back contact.
The absorber is a triple-cation perovskite with the composition Cs$_{0.05}$(Fa$_{83}$MA$_{17}$)$_{0.95}$PbI$_{83}$Br$_{17}$.
Moreover, PTAA is used as HTL, and C$_{60}$ as ETL.

In the considered device, the dominant recombination pathways included in our model are radiative recombination,\autocite{Roosbroeck1954Radiative} non-radiative Shockley-Read-Hall recombination,\autocite{Shockley1952, Hall1952} and interfacial recombination.
Their mathematical formulations are provided in Sec.\ \ref{supp:Electronic-Model} of the Electronic Supplementary Information (ESI).
For many perovskite solar cells, interfacial recombination constitutes a major limitation to device performance.\autocite{stolterfoht2019impact, Wolff2017Interface, Diekmann2021}
In our specific configuration, Auger-Meitner recombination has only a minor influence on the power-conversion efficiency and is therefore neglected.\autocite{Diekmann2021}

We consider both planar and nanotextured versions of this architecture (Fig.\ \ref{fig:1}b).
The planar configuration (left) serves as a reference, while the nanotextured version (right) introduces sinusoidal textures at the glass/ITO, ITO/HTL, and HTL/PVK interfaces.
Sinusoidal hexagonal nanotextures enabled a world-record power conversion efficiency for perovskite-silicon tandem solar cells between late 2021 and mid 2022.\autocite{tockhorn2022nano}
Earlier work on single-junction perovskite solar cells showed that the optical and electronic performance of sinusoidal nanotextures was superior to that of inverted pyramids and pillars.\autocite{tockhorn2020improved}
In those works, the perovskite layers were spin-coated, which is incompatible with state-of-the art pyramid textures that have been used for silicon solar cells for a long time.\autocite{haynos:1974}

To reduce the computational load, we perform 2D simulations, assuming the textures to be invariant along the out-of-plane direction.
The considered sinusoidal texture is given by
\begin{equation}
  y(x) =  h_{\text{T}} - h_{\text{T}}\cos\left(\frac{2\pi x}{w_{\text{T}}}\right),
\end{equation}
where $h_{\text{T}}$ is the variable texture height and $w_{\text{T}} = 750$\,nm the fixed period width, which is fixed for all texture heights.

We varied $h_{\text{T}}$ between $0$\,nm (planar) and $750$\,nm.
For the planar case, all layer thicknesses of the ETL ($30$\,nm), PVK ($400$\,nm), and HTL ($10$\,nm) match the experimentally measured values.\autocite{LeCorre2022,thiesbrummel2024ion}
For textured devices ($h_{\text{T}} >0$\,nm), we ensure that the total PVK volume remains constant across all configurations by adjusting the PVK thickness below the texture.
This assumption guarantees that any increase in absorption is not due to an excess of material but only due to its distribution.
As the texture height increases, the length of the PVK/HTL interface increases as well, while the ETL/PVK interface remains unchanged.

Finally, in Fig.\ \ref{fig:1}c, we illustrate how the photogeneration data is transferred to the electronic charge transport simulations:
First, the generation rate is computed on a finite element (FE) mesh (left panel) by solving the time-harmonic Maxwell equations.
This data is then interpolated onto uniform Cartesian grid points for the data export (middle panel), and mapped onto a finite volume (FV) mesh (right panel) used in the drift-diffusion simulations.
These simulations provide access to the total current density, spatial distributions of carrier concentrations, electric fields, and recombination rates.

As prototyping a series of different nanotextured cells is comparatively costly and does not readily yield physical insight into the electronic behaviour, we focus here on numerical simulations.

\subsubsection*{Influence of texturing on light absorption and carrier generation}
\begin{figure*}[ht]
  \centering
  \includegraphics[width=\textwidth]{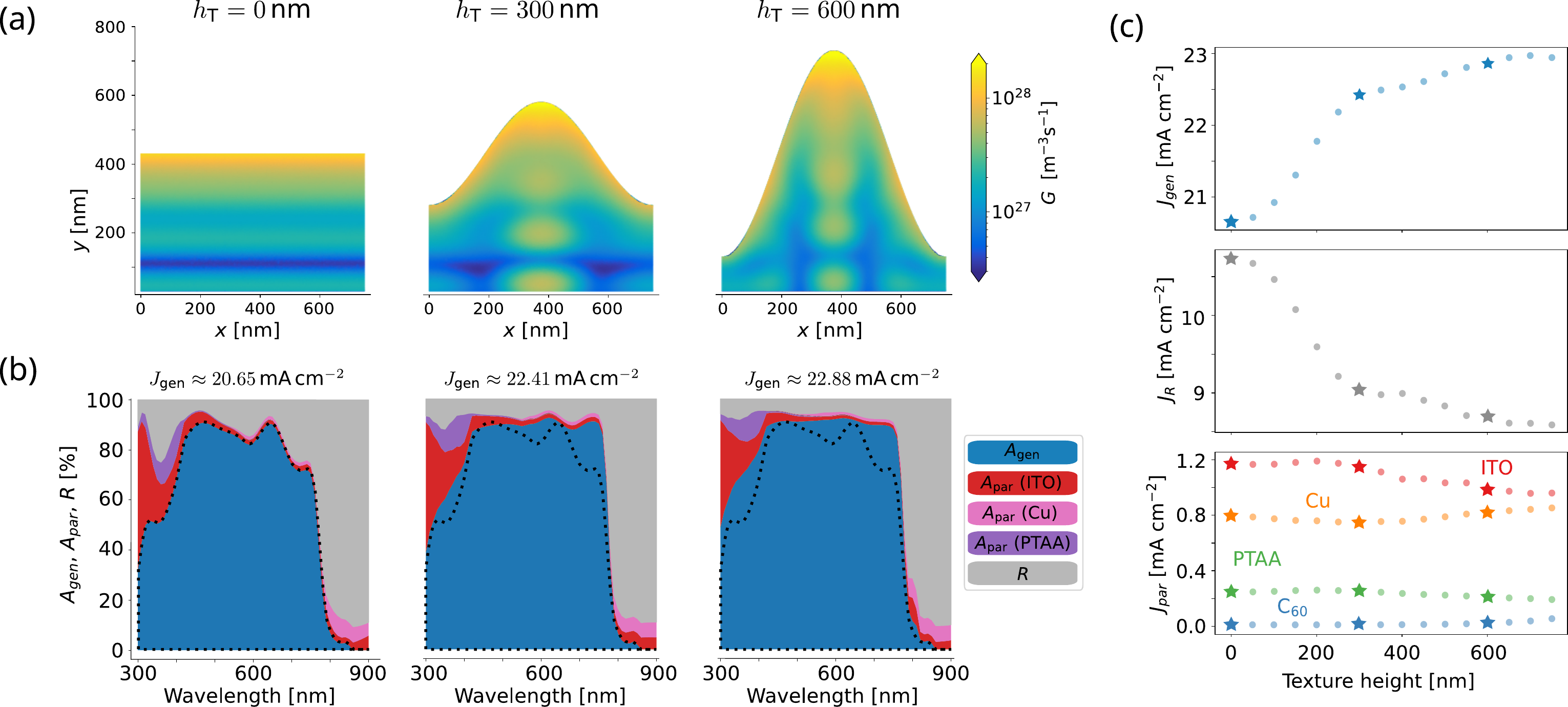}
  \caption{
    (a) The optical photogeneration rate $G$ plotted as a function of position in the perovskite layer for cells with no texture (left panel), $300$\,nm (mid panel), and $600$\,nm (right panel) nanotexture height.
    (b) Total spectral reflectance ($R$), parasitic absorptance ($A_\text{par}$ in PTAA, Cu and ITO) and absorptance ($A_{\text{gen}}$) for no texture (left panel), $300$\,nm texture (mid panel), and $600$\,nm texture (right panel).
    The black dashed line indicates the absorptance for the non textured case of the first panel.
    The maximum achievable short-circuit current density $J_{\text{gen}}$ calculated from the photogeneration rate within the perovskite absorber is stated above each panel, respectively.
    (c) Dependence of the maximum achievable short-circuit current density $J_{\text{gen}}$ (top panel), the reflective losses $J_{R}$ (mid panel), and the parasitic losses $J_{\text{par}}$ (bottom panel) on texture height.
    The stars indicate the three textures shown in (a) and (b).
  }
  \label{fig:2}
\end{figure*}

Solar cells are driven by the power of the incident light with its characteristic spectrum.
To compute the total reflectance, the absorptance in the different layers, and the photogeneration rate within the perovskite layer, we numerically solve the time-harmonic Maxwell equations in a scattering formulation.\autocite{Pomplun2007pssb}
The incident solar spectrum is discretised using monochromatic plane waves and weighted using the standardised AM1.5G reference spectrum.\autocite{NREL}
We calculated the spectral absorption density $\mathcal{A}_\text{gen}(\lambda,\mathbf{x})$ for the relevant spectral range from $\lambda_1 = 300$\,nm to $\lambda_2 = 900$\,nm with $10$\,nm step size.
The spectral absorption density quantifies where photons of a particular vacuum wavelength $\lambda$ are absorbed, and integrating it over all wavelengths yields the photogeneration rate
\begin{equation}\label{eq:photogeneration}
    G(\mathbf{x}) = \int_{\lambda_1}^{\lambda_2} \mathcal{A}_\text{gen}(\lambda,\mathbf{x}) \frac{\lambda}{hc} \,\d\lambda,
\end{equation}
which indicates where in the PVK layer the electron-hole pairs are generated.
In the integral, the absorption density is divided by the photon energy $hc \lambda^{-1}$, where $h$ is Planck's constant and $c$ is the speed of light \emph{in vacuo}.
Figure \ref{fig:2}a shows the photogeneration rate $G$ within the PVK layer for solar cells with no texture (left panel), an intermediate texture height of $300$\,nm (mid panel), and a large texture of $600$\,nm (right panel), all of which will be used in the electronic simulations in the next step.

Integrating $\mathcal{A}_\text{gen}$ over the total perovskite area $\Omega_{\text{PVK}}$ gives the spectral absorptance
\begin{equation}\label{eq:absorptance}
    A_\text{gen}(\lambda) = \int_{\Omega_{\text{PVK}}}\mathcal{A}_\text{gen}(\lambda,\mathbf{x})\d\mathbf{x}.
\end{equation}
Figure \ref{fig:2}b shows the spectral absorptance $A_\text{gen}$ within the PVK layer (blue), the parasitic absorptance $A_\text{par}$ in all remaining layers (red, purple, and pink) and reflectance $R$ (grey), in the relevant spectral range, for the three device geometries discussed above.
As known from previous work,\autocite{Duote18,tockhorn2020improved} the texture reduces the total reflectance and thus leads to more absorption in the PVK layer increasing charge carrier generation.
Further integrating either the absorptance over all wavelengths or the photogeneration rate over the area of the PVK layer $\Omega_{\text{PVK}}$, one obtains the maximum achievable short-circuit current density
\begin{equation}\label{eq:J_gen}
    J_{\text{gen}} = \frac{q}{w_{\text{T}}}\int_{\lambda_1}^{\lambda_2} A_{\text{gen}}(\lambda)\frac{\lambda}{hc}\d\lambda =\frac{q}{w_{\text{T}}}  \int_{\Omega_{\text{PVK}}}  G(\mathbf{x}) \, \d\mathbf{x},
\end{equation}
where $q$ is the elementary charge and $w_{\text{T}}$ is the texture width.
The maximum achievable short-circuit current density $J_{\text{gen}}$ serves as an upper bound to the short-circuit current density $J_{\text{SC}}$, since recombination prevents all generated electron–hole pairs from being collected at the contacts.
How the texture height affects recombination losses is discussed later, for now it is enough to state that these losses are only small compared to gains in $J_{\text{gen}}$.
Therefore, higher $J_{\text{gen}}$ current densities lead directly to higher $J_{\text{SC}}$ current densities.
Figure \ref{fig:2}c (top panel) shows that $J_{\text{gen}}$ strictly increases with texture height.
Thus, $J_{\text{SC}}$ gains are directly related to the improved optical properties of the textured devices.

To assess how much current density is lost via reflection and parasitic absorption, equivalent current densities $J_{R}$ and $J_i$ can be computed, where $A_\text{gen}$ in (\ref{eq:J_gen}) is replaced by $R(\lambda)$ or the parasitic absorption in the $i$-th layer.
Figure \ref{fig:2}c (middle panel) shows the  current density for the reflective losses $J_{R}$, which drops quickly with increasing texture height.
Figure \ref{fig:2}c (right panel) shows $J_\text{par}$ for the non-PVK layers of the device stack.
Most notably, the losses in the ITO layer decrease slightly for higher textures.
While the height of the ITO stays fixed, the texture increases the surface enhancement factor, effectively reducing its thickness for the incoming scattered light reducing the absorption in this layer.
The same effect applies to the PTAA.
Layers below the texture show a slight increase in their generated losses, due to an overall reduction in reflectance for all wavelengths.

Next, we export the results from the optical simulations to the electronic simulations by incorporating the photogeneration rates $G$ (Fig.\ \ref{fig:2}a) into the charge transport model.
More precisely, this data serves as source term in the electron ($\alpha = \electrons$) and hole ($\alpha = \holes$) drift-diffusion equations used in the electronic simulations,
\begin{equation}
   z_\alpha q \partial_t n_\alpha + \nabla\cdot \mathbf{j}_\alpha =  z_\alpha q   \left[G(\mathbf{x}) - R( n_\electrons, n_\holes) \right],
\end{equation}
explained in more detail in Sec.\ \ref{supp:Electronic-Model} (ESI).
We simulate the charge transport of electrons and holes in the ETL, PVK layer and HTL for different texture heights (Fig.~\ref{fig:3}a, top) while also accounting for mobile ion vacancies in the PVK layer.

\subsubsection*{Behaviour of simulated PCE, $\boldsymbol{J}_{\text{SC}}$, $\boldsymbol{V}_{\text{OC}}$, and FF}

In the following, we present simulation results, including current density-voltage ($J$-$V$) curves, open-circuit voltages $V_{\text{OC}}$, short-circuit current densities $J_{\text{SC}}$, power conversion efficiencies (PCEs), and fill factors (FFs).

\begin{figure*}[ht]
  \centering
  \includegraphics[width=\textwidth]{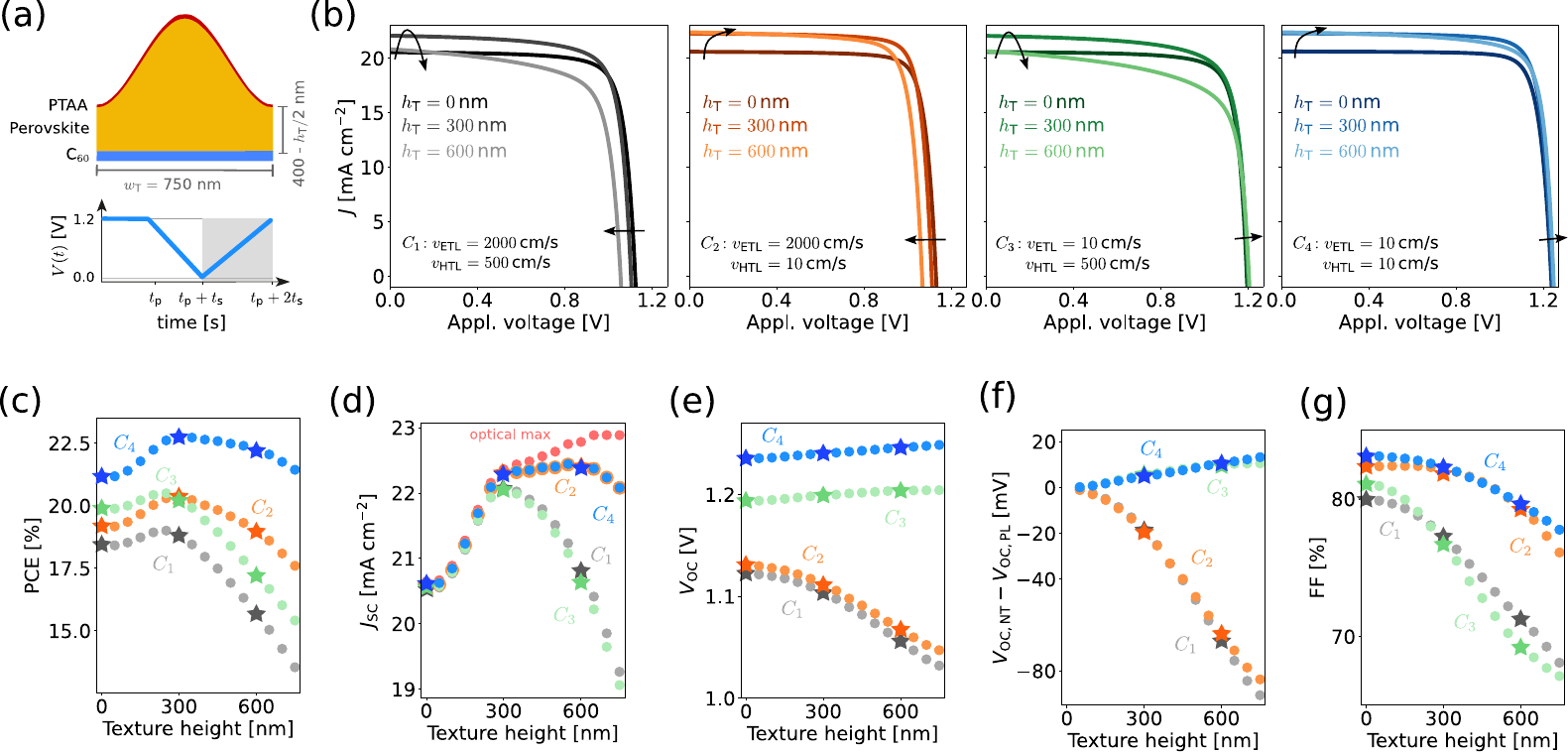}
  \caption{
      Calculated performance metrics for the studied single-junction solar cell by solving the drift-diffusion charge transport model with a given optical photogeneration rate, computed from Maxwell's equations.
      (a) The device geometry (top) considered for the electronic simulations, consisting of the electron transport layer (C$_{60}$), the ``83-17 triple cation'' perovskite material layer, and the hole transport layer (PTAA).
      The thickness of the perovskite layer varies with the texture height $h_{\text{T}}$, while the texture width $w_{\text{T}}$ stays the same.
      Moreover, the considered measurement protocol (bottom) includes a preconditioning step, a backward and a forward scan.
      The following simulation results correspond to the grey shaded forward scan.
      (b) Simulated forward current-voltage ($J$-$V$) curves for cases $C_1$ to $C_4$, where the surface recombination velocities $v_{\text{HTL}}$ and $v_{\text{ETL}}$ are varied.
      Brighter colours indicate greater texture height, with arrows showing the direction of increasing texture height.
      Furthermore, we show for all test cases $C_1$ (grey), $C_2$ (orange), $C_3$ (green), and $C_4$ (blue), the impact of texture height on
      (c) the power conversion efficiency (PCE),
      (d) the short-circuit current density $J_{\text{SC}}$,
      (e) the open-circuit voltage $V_{\text{OC}}$,
      (f) the difference in $V_{\text{OC}}$ between planar (PL) and textured systems (NT), and
      (g) the fill factor (FF).
      The stars indicate the three texture heights shown in (b).
    }
  \label{fig:3}
\end{figure*}

The simulated $J$-$V$ characteristics are obtained using a fast hysteresis measurement technique, designed to assess ionic redistribution losses at different scan speeds.\autocite{LeCorre2022}
In this procedure, illustrated in Fig.\ \ref{fig:3}a (bottom), the devices are held at a constant voltage near the open-circuit voltage ($V_{\text{max}} \geq V_{\text{OC}}$) for $t_{\text{p}}$ seconds, followed by backward and forward voltage scans, each lasting $t_{\text{s}} = V_{\text{max}}/f$ seconds, where $f$ denotes the scan rate.

All simulations use a fast scan rate of $f = 10^{3}$\,V$\cdot$s$^{-1}$.
At this rate, ionic motion is negligible, as shown in Fig.~\ref{supp:figs5} and Fig.~\ref{supp:figs10} (ESI), where the vacancy density remains constant across different applied voltages.
Throughout our study, the average vacancy concentration is fixed to $6.0 \times 10^{22}$\,m$^{-3}$ within the perovskite layer, following values reported in the literature.\autocite{LeCorre2022}
As ionic motion is suppressed at this scan rate, the device operates without hysteresis.\autocite{thiesbrummel2024ion, LeCorre2022}
We therefore restrict our analysis to the forward scan.

Next, we extend the planar baseline to textured systems and investigate four scenarios with varying surface recombination velocities at the carrier transport layers.
With increasing texture height, the PVK/HTL interface becomes longer, while the ETL/PVK interface remains unchanged.
Case $C_1$ corresponds to the parameter set used by Thiesbrummel \emph{et al.}\autocite{thiesbrummel2024ion}. In the subsequent cases, we selectively reduce the surface recombination velocity at the HTL interface ($C_2$), at the ETL interface ($C_3$), or at both interfaces simultaneously ($C_4$).
In other words, we have the test cases:
\begin{itemize}
  \item[$C_1$:] $v_\text{ETL}=2000$\cms,  $v_\text{HTL}=500$\cms (ref.\ configuration\citenum{thiesbrummel2024ion})\\[-3.6ex]
  \item[$C_2$:] $v_\text{ETL}=2000$\cms, $v_\text{HTL}=\phantom{0}10$\cms\\[-3.6ex]
  \item[$C_3$:] $v_\text{ETL}=\phantom{00}10$\cms, $v_\text{HTL}=500$\cms\\[-3.6ex]
  \item[$C_4$:] $v_\text{ETL}=\phantom{00}10$\cms, $v_\text{HTL}=\phantom{0}10$\cms
\end{itemize}

Figure~\ref{fig:3}b displays the $J$-$V$ curves for cases $C_1$ -- $C_4$. Within each subfigure, brighter colours indicate larger texture heights, and arrows mark the direction of increasing texture height.
In the reference case $C_1$, the short-circuit current density $J_{\text{SC}}$ reaches its maximum at a texture height of $h_\text{T} = 300$\,nm, but decreases again for larger values of $h_\text{T}$.
The open-circuit voltage $V_{\text{OC}}$, however, decreases monotonically with texture height.
Reducing the surface recombination velocity $v_{\text{HTL}}$ at the HTL ($C_2$) has a beneficial effect on the $J_{\text{SC}}$:
It increases consistently with texture height, while the behaviour of $V_{\text{OC}}$ remains essentially unchanged compared with the reference case $C_1$.
In contrast, reducing the surface recombination velocity at the ETL ($C_3$) leads to the opposite trend:
the qualitative behaviour of $J_{\text{SC}}$ (initial rise followed by a decline) is largely preserved, but $V_{\text{OC}}$ now increases with texture height.
When both surface recombination velocities are reduced ($C_4$), the advantages of both modifications ($C_2$) and ($C_3$) are combined, yielding an increase in both $J_{\text{SC}}$ and $V_{\text{OC}}$ for larger texture heights.

We analyse this behaviour quantitatively in the second row of Fig.~\ref{fig:3}c-g.
We show the power conversion efficiency (PCE), the short-circuit current density $\JSC$, the open-circuit voltage $\VOC$, and the fill factor (FF) as functions of texture height for all test cases $C_1$ to $C_4$.
All four configurations have maximal power conversion efficiencies (PCE, Fig.~\ref{fig:3}c) between $h_\text{T} = 250$\,nm and $h_\text{T} = 300$\,nm.

\newpage

For all texture heights, the PCE of the reference case $C_1$ remains below that of the low-surface-recombination case $C_4$, while the mixed cases $C_2$ and $C_3$ yield PCE values between these two extremes.
The origin of the PCE maximum near $h_\text{T} = 250$\,nm or $h_\text{T} = 300$\,nm can be attributed partly to the behaviour of the short-circuit current density $J_{\text{SC}}$ (Fig.~\ref{fig:3}d), which deviates substantially from the maximum achievable short-circuit current density (red dots) beyond this texture height.
The trends in $J_{\text{SC}}$ with texture height for $C_1$ and $C_3$ (high $v_{\text{HTL}}$) behave similarly, as those do for $C_2$ and $C_4$ (low $v_{\text{HTL}}$), with the latter group staying much closer to the optical ideal.

Figure~\ref{fig:3}e shows the dependence of the open-circuit voltage $V_{\text{OC}}$ on texture height.
Reducing $v_{\text{HTL}}$ ($C_2$) has only a minor impact on $V_{\text{OC}}$, which continues to decrease with increasing texture height much like in the reference case ($C_1$).
In contrast, reducing $v_{\text{ETL}}$ significantly improves the open-circuit voltage ($C_3$ and $C_4$) and leads to an increasing $V_{\text{OC}}$ with increasing with texture height.

Although the sinusoidal texture increases the HTL interface area, changes in $v_{\text{HTL}}$ ($C_2$) only weakly affect the $V_{\text{OC}}$.
Instead, the dominant sensitivity arises from the \emph{unchanged} ETL interface, where reducing $v_{\text{ETL}}$ ($C_3$, $C_4$) markedly improves the $V_{\text{OC}}$ and even reverses its trend with texture height.
This behaviour indicates that ETL surface recombination mainly governs the device near open-circuit conditions, whereas the enlarged HTL interface has a stronger impact only at lower voltages.
Interestingly, the absolute difference between the planar and textured open-circuit voltages behaves the same for $C_1$ and $C_2$ (high $v_{\text{ETL}}$), and $C_3$ and $C_4$ (low $v_{\text{ETL}}$), as can be seen in Fig.~\ref{fig:3}f.
Finally, the fill factor (FF) shown in Fig.~\ref{fig:3}g decreases with increasing texture height for all four cases.

\begin{table}[h]
  \centering
    \caption{Simulated current-voltage parameters for the planar system as well as for the textured systems with the highest power conversion efficiencies (PCEs).
    For the cases $C_1$ and $C_3$, the highest PCE is reached for $h_{\text{T}} = 250$\,nm, while in the cases $C_2$ and $C_4$ for $h_{\text{T}} = 300$\,nm
    }
    \label{tbl:Tab1}
    \begin{tabular*}{0.78\textwidth}{@{\extracolsep{\fill}}llllll}
      \hline
        & & PCE [\%] & $J_{\text{SC}}$ [mA$\cdot$cm$^{-2}$]  & $V_{\text{OC}}$ [V] & FF [\%] \\[0.3ex]
      \hline\\[-1.4ex]
      $C_1$ & planar &  18.4 &  20.5 & 1.123  & 79.9 \\[0.5ex]
      & textured &  19.0 &  21.9 & 1.110  & 78.1 \\[1.3ex]
      $C_2$ & planar &  19.2 &  20.6 & 1.131  & 82.4 \\[0.5ex]
      & textured &  20.3 &  22.3 & 1.111  & 82.2 \\[1.3ex]
      $C_3$ & planar &  19.9 &  20.6 & 1.194  & 81.1 \\[0.5ex]
      & textured &  20.5 &  21.9 & 1.198  & 78.0 \\[1.3ex]
      $C_4$ & planar &  21.2 &  20.6 & 1.235  & 83.1 \\[0.5ex]
      & textured &  22.7 &  22.3 & 1.240  & 82.3 \\
      \hline
    \end{tabular*}
\end{table}

Now let us put these results into perspective.
Tockhorn \emph{et al.}\ \autocite{tockhorn2020improved} reported results on a textured single-junction perovskite solar cell with 'cos-' texture with $h_\text{T} = 220$\,nm texture height:
For the forward scan, $0.6$\,\%$_{\text{abs}}$ absolute efficiency gain, $1$\,mA\,cm$^{-2}$ increase in $J_{\text{SC}}$, $20$\,mV increase in $V_{\text{OC}}$, and $1$\,\%$_{\text{abs}}$ loss in the fill factor were observed with respect to a planar reference.

Our opto-electronic simulations exhibit similar trends to these experimental observations as summarised in Tab.\ \ref{tbl:Tab1}.
Specifically, for the reference configuration\autocite{thiesbrummel2024ion} $C_1$, the PCE is maximally increased by $0.6$ \%$_{\text{abs}}$ for a texture height of $250$\,nm compared to a planar device.
Significant increases greater than $1.0$ \%$_{\text{abs}}$ in the PCE occur for the two test cases, where $v_{\text{HTL}}$ ($C_2$ and $C_4$) is decreased.
In all surface recombination velocity configurations the increase in $\JSC$ is higher than $1.3$\,mA\,cm$^{-2}$.
The losses in the fill factor behave roughly the same.
In case of high $v_{\text{ETL}}$ ($C_1$ and $C_2$), the losses in $\VOC$ are $ > 12$\,mV, while in case of low $v_{\text{ETL}}$ ($C_3$ and $C_4$), the increase in $\VOC$ are $ \approx 4-5$\,mV.

The PCE is proportional to $\JSC\cdot\VOC\cdot\text{FF}$.
For all test cases, the relative changes in $J_{\text{SC}}$ are approximately an order of magnitude larger than those in $V_{\text{OC}}$ and FF.
Therefore, the enhancement in PCE arises primarily from the increases in short-circuit current density.

Finally, to estimate how much of the observed increase in $V_{\text{OC}}$ can be attributed purely to the optical enhancement in $J_{\text{SC}}$, we use the classical Shockley diode relation for p-n junctions, \autocite{Wuerfel2016} which gives
\begin{align} \label{eq:ShockleyDiode}
  \Delta V_{\text{OC}} \approx \frac{k_{\text{B}} T}{q} \ln \delta_{\text{SC}},
\end{align}
where $\delta_{\text{SC}} = J_{\text{SC, NT}} / J_{\text{SC, PL}}$ is the ratio of short-circuit current densities for the textured (NT) and planar (PL) systems, $k_{\text{B}}$ is the Boltzmann constant, and $T$ is the temperature.
This expression assumes that the dark saturation current density $J_0$ is much smaller than $J_{\text{SC}}$, that $J_0$ remains constant for different texture heights, and that the ideality factor $n$ is set to unity.

From Tab.~\ref{tbl:Tab1}, we find $\delta_{\text{SC}} \approx 1.07$ for $C_3$ and $\delta_{\text{SC}} \approx 1.08$ for $C_4$, yielding expected increases in $V_{\text{OC}}$ of $\Delta V_{\text{OC}} \approx 1.75$\,mV ($C_3$) and $\Delta V_{\text{OC}} \approx 2.01$\,mV ($C_4$).
By comparison, the simulated $V_{\text{OC}}$ increase is roughly two to three times larger than this estimate, indicating that mechanisms, beyond the increase in $J_{\text{SC}}$, play a role.

In summary, our simulations show that reducing $v_{\text{HTL}}$ keeps $J_{\text{SC}}$ closer to the optical ideal as the texture height increases, which impacts the PCE the most.
In contrast, lowering $v_{\text{ETL}}$ can even counteract the detrimental impact of larger texture heights on $V_{\text{OC}}$. Across all configurations, the maximum PCE is attained near $h_\text{T} = 300$\,nm.
To investigate the origin of these observations, we next examine the recombination processes in the simulated systems.

\begin{figure*}[ht]
  \centering
  \includegraphics[width=\textwidth]{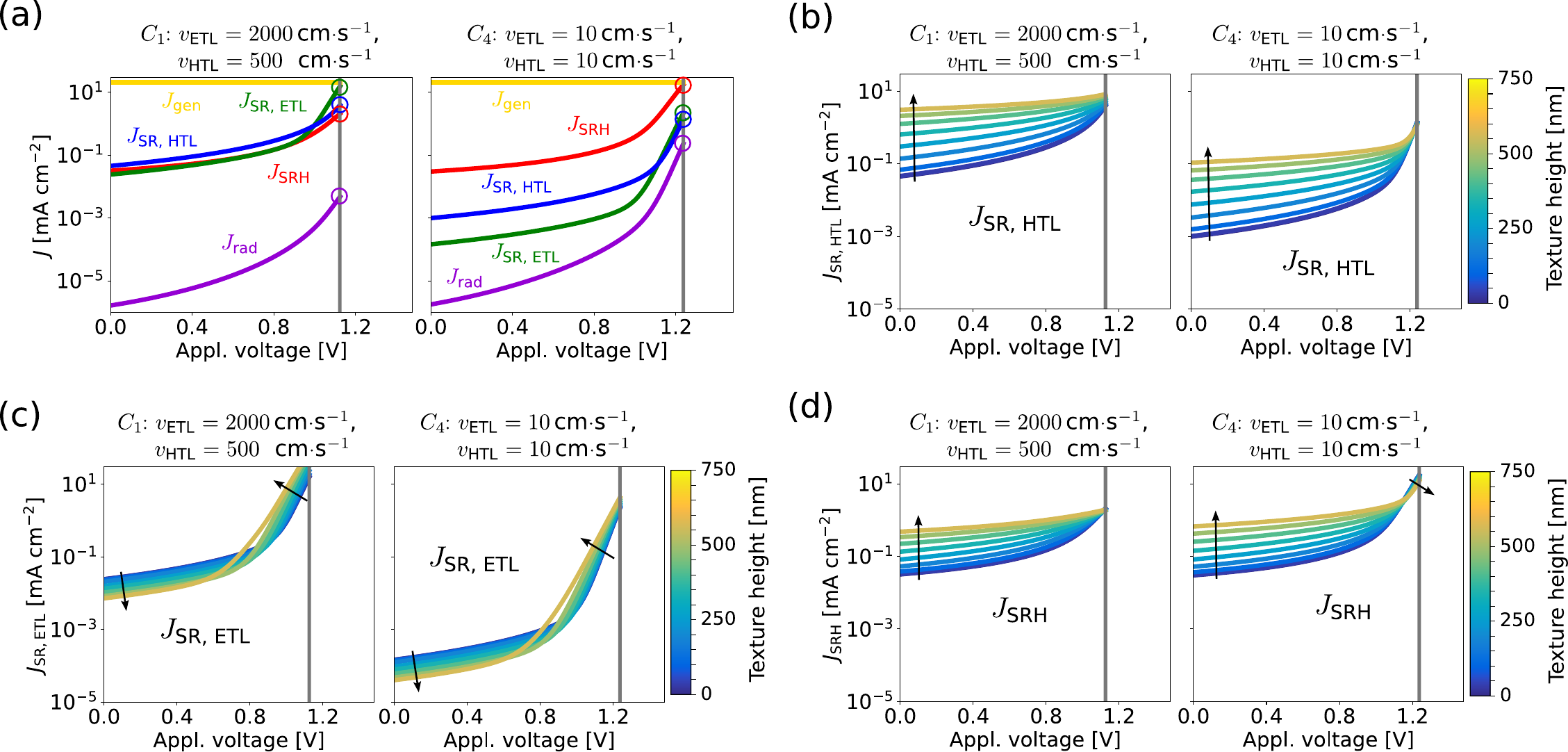}
  \caption{
    Simulated voltage-dependent recombination current densities for planar and textured perovskite devices for the reference high-surface-recombination case $C_1$ (left) and the low-surface-recombination case $C_4$ (right) from drift-diffusion calculations, with the photogeneration rate obtained from Maxwell's equations.
    (a) Recombination current densities for the planar systems. We included radiative $J_{\text{rad}}$ and surface recombination at the PVK/HTL $J_{\text{SR, HTL}}$ and ETL/PVK interfaces $J_{\text{SR, ETL}}$ as well as Shockley-Read-Hall (SRH) $J_{\text{SRH}}$ recombination.
    These integrated recombination rates are compared to the generation current $J_{\text{gen}}$.
    Furthermore, we show for both recombination velocity test cases
    (b) $J_{\text{SR, HTL}}$,
    (c) $J_{\text{SR, ETL}}$, and
    (d) $J_{\text{SRH}}$ for varying texture height.
    The grey vertical line indicates the open-circuit voltage of the planar configuration.
    The arrows indicate the direction of increasing texture height. Colour coding indicates device morphology: blue corresponds to planar devices and yellow to strongly textured devices.
    }
  \label{fig:4}
\end{figure*}

\subsubsection*{Impact of texturing on recombination currents}

In this section, we identify the dominant recombination mechanisms across different voltage-bias regimes and analyse their influence on $V_{\text{OC}}$, $J_{\text{SC}}$, and consequently on the device efficiency.
The individual recombination rates depend on the electron and hole densities, as detailed in the ESI (Section~\ref{supp:Electronic-Model}).
Throughout the forward scan, we first compute the carrier densities and subsequently evaluate the recombination rates in a post-processing step.
For clarity, we restrict our discussion to the representative cases $C_1$ and $C_4$:
\emph{(i)} the reference configuration $C_1$, characterised by high surface recombination velocities $v_{\text{ETL}}$ and $v_{\text{HTL}}$, and
\emph{(ii)} the low-surface-recombination configuration $C_4$, in which both $v_{\text{ETL}}$ and $v_{\text{HTL}}$ are strongly reduced.
The intermediate cases $C_2$ and $C_3$ behave qualitatively like $C_1$ and $C_4$.

Figure~\ref{fig:4}a displays the recombination current densities for planar devices in these two configurations for the forward scan.
We show the radiative recombination current density $J_{\text{rad}}$ (purple), the surface recombination current densities at the HTL interface $J_{\text{SR, HTL}}$ (blue) and at the ETL/PVK interface $J_{\text{SR, ETL}}$ (green), and the Shockley-Read-Hall (SRH) recombination current density $J_{\text{SRH}}$ (red).
For reference, the generation current density $J_{\text{gen}}$ (yellow) is also included.

As expected from Thiesbrummel \emph{et al.},\autocite{thiesbrummel2024ion} the reference case $C_1$ exhibits dominant recombination at the ETL/PVK interface, as seen in Fig.~\ref{fig:4}a (left).
In contrast, when both surface recombination velocities are strongly reduced ($C_4$), this suppression shifts the dominant loss mechanism to SRH recombination, as shown in Fig.~\ref{fig:4}a (right).

To analyse how nanotexturing alters these recombination rates, Figures~\ref{fig:4}b-d show the texturing-induced changes in $J_{\text{SR, HTL}}$, $J_{\text{SR, ETL}}$, and $J_{\text{SRH}}$, respectively.
The colour code spans from blue (planar device) to yellow (strong texture), and arrows indicate increasing texture height.
Radiative recombination is orders of magnitudes smaller than the other rates and largely unaffected by texturing, as shown in Fig.\ \ref{supp:figs1} (ESI).
Hence, it is not displayed here.

While the absolute magnitudes of the recombination currents shift depending on the chosen surface recombination velocities $v_{\text{HTL}}$, $v_{\text{ETL}}$, their qualitative dependence on texture height remains robust.
For this reason, we focus on the following trends, where the first three trends are independent of the chosen surface recombination velocities:
\begin{enumerate}
    \item[(T1)] {$J_{\text{SC}}$ conditions ($V = 0$\,V):}
    Both $J_{\text{SR, HTL}}$ and $J_{\text{SRH}}$ increase with texture height.\\[-3.9ex]
    \item[(T2)] {$J_{\text{SC}}$ conditions ($V = 0$\,V):}
    $J_{\text{SR, ETL}}$ decreases with texture height.\\[-3.9ex]
    \item[(T3)] {Near $V_{\text{OC}}$ conditions:}
    $J_{\text{SR, ETL}}$ increases with texture height.\\[-3.9ex]
    \item[(T4)] {Near $V_{\text{OC}}$ conditions (only for low $v_{\text{ETL}}$):}
    $J_{\text{SRH}}$ decreases with texture height.
\end{enumerate}

The final trend (T4) is special because it appears only when the ETL surface recombination velocity is low, so that SRH recombination in the bulk becomes the dominant loss mechanism.
In these bulk-recombination-dominated solar cells, the open-circuit voltage increases with texture height.
Trend (T1) explains the strong impact of texturing on $J_{\text{SC}}$ via increased HTL-side recombination, whereas a combination of (T3) and (T4) explains why texturing affects $V_{\text{OC}}$ through increased ETL-side recombination.

For the intermediate cases $C_2$ and $C_3$, the recombination current densities
$J_{\text{SR, HTL}}$ (Fig.\ \ref{supp:figs2}),
$J_{\text{SR, ETL}}$ (Fig.\ \ref{supp:figs3}) and
$J_{\text{SRH}}$ (Fig.\ \ref{supp:figs4})
are shown in the ESI.
These figures confirm that trends (T1) to (T3) also apply to these mixed cases.
Trend (T4), however, appears only when $v_{\text{ETL}}$ is sufficiently reduced for SRH recombination to become the dominant loss mechanism, as can be seen from Fig.\ \ref{supp:figs4} (ESI).

These observations clarify the mechanisms behind the observed changes in $J_{\text{SC}}$, $V_{\text{OC}}$, and PCE:
Surface recombination at the HTL primarily affects $J_{\text{SC}}$, while recombination at the ETL mainly influences $V_{\text{OC}}$.
The physical origin of trends (T1) to (T4) becomes apparent when examining the electric-field and carrier density configurations, which we analyse in the following subsection.

\subsubsection*{Effect of texturing on electric field and charge carriers}

\begin{figure*}[ht]
  \centering
  \includegraphics[width=\textwidth]{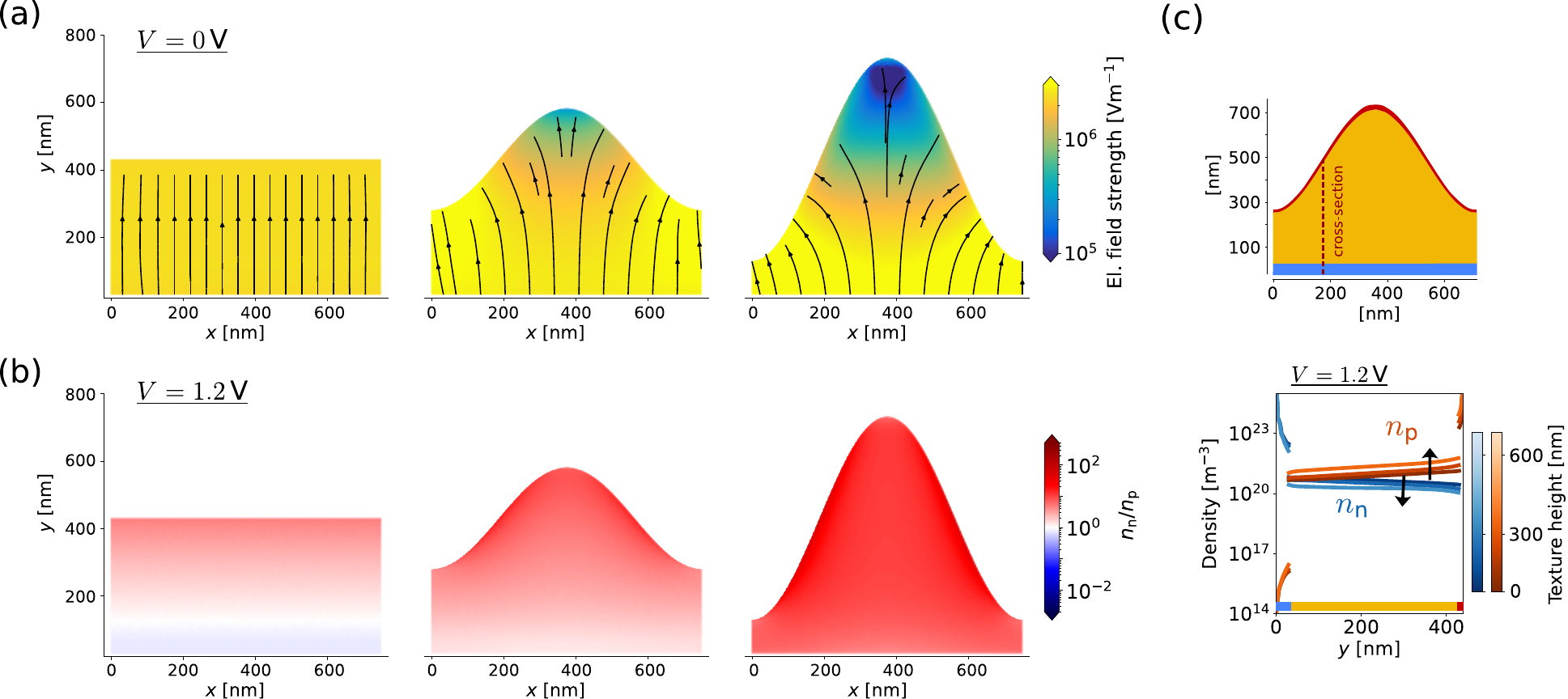}
  \caption{
    (a) Electric field for three texture heights $h_{\text{T}} = 0, 300, 600$\,nm for $V = 0$\,V applied voltage.
    The colour and the stream plot indicate the strength $\Vert - \nabla \psi \Vert_2 $ and the direction of the electric field, respectively.
    (b) The corresponding ratio between hole and electron density $n_\holes/n_\electrons$ for $V = 1.2$\,V applied voltage.
    (c) 2D device geometry with the vertical cross-section indicated (top), along which the carrier densities (bottom) are extracted.
    More precisely, we see one-dimensional profiles of the electron (blue) and hole densities (red) at an applied voltage $V = 1.2$\,V for varying texture height.
    In the density plot, brighter colours indicate greater texture height, with arrows showing the direction of increasing texture height.
    All results correspond to the low-recombination case $C_4$.
    }
  \label{fig:5}
\end{figure*}

Figure~\ref{fig:5} shows the electric field and carrier densities of electrons and holes at different applied voltages for the low-surface-recombination case $C_4$ (low $v_{\text{ETL}}$, low $v_{\text{HTL}}$).
For this velocity combination, all four previously discussed trends (T1)-(T4) are observed and can be explained.

Figure~\ref{fig:5}a shows the electric field at short-circuit condition ($V = 0$\,V).
The electric field vectors, indicated by stream plots, point in the direction in which holes are driven by the field.
Moreover, Fig.~\ref{fig:5}b visualizes the ratio between hole and electron density, while Fig.~\ref{fig:5}c shows one-dimensional cross sections of electron and hole densities near open-circuit voltage ($V = 1.2$\,V).

At $V = 0$\,V, the electric field is homogeneous in the planar configuration, but nanotexturing redistributes it:
The field strengthens in the valleys and weakens at the peaks.
Therefore, textured devices experience enhanced charge separation in the valleys, whereas the reduced field at the peaks leads to local carrier accumulation and thus increased recombination.
This explains why HTL surface recombination and SRH recombination increase in trend (T1), whereas ETL surface recombination decreases at the beginning of the forward scan in trend (T2).

At higher applied voltages near open-circuit ($V = 1.2$\,V), the internal electric field weakens and drift becomes negligible (ESI, Fig.\ \ref{supp:figs8}).
We therefore analyse the carrier densities next to understand the trends (T3) and (T4).
From Fig.\ \ref{fig:5}b, we find that the perovskite layer contains significantly more holes than electrons for textured systems as the device approaches open-circuit conditions.
The inequality $n_\holes > n_\electrons$ holds for all larger texture heights and across the entire perovskite layer.
This results from the fact that increasing the texture height also increases the effective PVK/HTL interface length, increasing hole injection from the HTL.
Consequently, more holes reach the ETL, and surface recombination at the ETL increases with texturing [trend (T3)].


Hou \emph{et al.}\autocite{Hou_2020} speculated that extended drift-dominated regions near the PVK/HTL interface were responsible for enhanced $V_{\text{OC}}$ in textured perovskite solar cells.
In contrast, we find that drift near the peaks is always reduced.
Therefore, another mechanism must be responsible for the enhanced $\VOC$ observed in nanotextured devices.

To understand trend (T4), the reduction of SRH recombination near open-circuit voltage, and by that the enhanced $\VOC$ values, we examine Fig.\ \ref{fig:5}c, which shows cross sections of the electron and hole densities.
As shown in the ESI (Fig.\ \ref{supp:figs7}), the carrier densities vary minimally along the $x$-direction near $V \approx \VOC$.
Therefore, in Fig.\ \ref{fig:5}c we focus on one-dimensional cross-sections, which correspond to $x \approx 187$\,nm, where the combined thickness of the ETL, perovskite (PVK) layer, and HTL is $y \approx 440$\,nm for all texture heights.

A direct consequence of the rational form of the steady-state Shockley-Read-Hall expression is that the ratio $n_\holes/n_\electrons$ influences the recombination rate, even when the product $n_\electrons n_\holes$ remains constant.
For a deep defect, the SRH recombination rate is maximised for fixed $n_\electrons n_\holes$, when $\tau_\holes n_\electrons = \tau_\electrons n_\holes$.\autocite{Huepkes2022}
In our setup, the carrier lifetimes are equal.\autocite{LeCorre2022,thiesbrummel2024ion,Diekmann2021}
Thus, SRH recombination is highest when $n_\holes / n_\electrons \approx 1$.
Figure~\ref{fig:5}c (bottom) shows an increasing imbalance between electron and hole densities with increasing texture height.
Specifically, we have $n_\holes > n_\electrons$ (Fig.~\ref{fig:5}b), which directly decreases the SRH recombination rate for textured systems.

The quasi Fermi level splitting (QFLS), which is given by the energy difference between the electron and hole quasi Fermi levels, is a key quantity in determining the maximum achievable $V_{\text{OC}}$.
The QFLS $\Delta E_{\text{F}}$ can be directly related to the product of electron and hole densities via
\begin{equation} \label{eq:QFL}
  \Delta E_{\text{F}} = k_B T \ln \left( \frac{n_\electrons n_\holes}{n_i^2} \right),
\end{equation}
where denotes $n_i$ the intrinsic carrier density.\autocite{Wuerfel2016}
It is well-established that in perovskite-based solar cells, the QFLS does not necessarily equal the open-circuit voltage $V_{\text{OC}}$ when interfacial energy offsets are present.\autocite{stolterfoht2019impact, caprioglio2019relation}
The QFLS represents a theoretical upper limit for $V_{\text{OC}}$.
In the ideal case of perfectly selective transport layers, we expect $\Delta E_{\text{F}} \approx qV_{\text{OC}}$.
For a bias near open-circuit ($V = 1.2$\,V), the integral average over the perovskite layer of the electron and hole densities $\overline{n_\electrons n_\holes}$ increases with texture height (see ESI, Fig.~\ref{supp:figs7}),
\begin{align*}
    \sqrt{\overline{n_\electrons n_\holes}}
    &\approx  5.88 \times 10^{20}\,\text{m}^{-3},
    \quad \text{for} \;\, h_\text{T} = \phantom{00}0\,\text{nm},\\
    \sqrt{\overline{n_\electrons n_\holes}}
    &\approx  7.30 \times 10^{20}\,\text{m}^{-3},
    \quad \text{for} \;\, h_\text{T} = 600\,\text{nm}.
\end{align*}
Consequently, the logarithmic term in Eq.~\eqref{eq:QFL} increases, lifting the upper limit for $V_{\text{OC}}$ with growing texture height.

Finally, the electric field and the carrier densities show the same qualitative behaviour for the reference surface recombination case $C_1$, as visualized in Fig.\ \ref{supp:figs9} (ESI).
The surface-recombination velocities modify only the magnitude of the resulting recombination rates, not the underlying qualitative trends (T1) to (T3) coming from texturing.

\subsection*{Conclusions}

In this theoretical work, we build on the well-studied planar single-junction perovskite solar cell setup\autocite{thiesbrummel2024ion,Diekmann2021, LeCorre2022} to investigate how sinusoidal nanotextures between material layers affect device electronics.
Multi-dimensional optical simulations were used to calculate the photogeneration rate, which served as input for subsequent electronic simulations.
By analysing recombination rates, electric fields, and carrier distributions, we quantified how nanostructures influence the electronic performance of perovskite solar cells under different surface recombination configurations.

Texturing redistributes the electric field, strengthening it in valleys and weakening it at peaks, thereby affecting carrier accumulation and recombination dynamics.
Across all recombination configurations, texturing improved the power conversion efficiency, with the highest values at a texture height of around $300$\,nm.
Surface recombination plays a central role when texturing:
The responses of $\JSC$ and $\VOC$ depend sensitively on the recombination velocities at the transport layers.
Reducing the HTL recombination velocity helps to maintain the optical $\JSC$ enhancement at larger texture heights, as texturing primarily increases HTL surface recombination at lower voltages.
In contrast, lowering the ETL recombination velocity increases $\VOC$ beyond what is expected from improved light absorption alone, because SRH recombination near $\VOC$ conditions decreases with increasing texture height due to an increased carrier imbalance ($n_\holes > n_\electrons$).

These findings provide clear design guidelines for high-efficiency nanotextured perovskite solar cells: effective passivation of the flat ETL interface is crucial to unlock $\VOC$ gains, while passivation of the textured HTL interface is essential to maximize $\JSC$ improvements.
In addition, such topology optimization may lead to efficiency gains not only in solar cells but also in light-emitting diodes and photodetectors.

\subsection*{Methods}

\subsubsection*{Optical model and simulation}
For the optical simulations we numerically solve the time-harmonic wave equation as derived from Maxwell's equations, formulated as a scattering problem.
We use the finite element method (FEM) as implemented in the software \texttt{JCMsuite}.\autocite{Pomplun2007pssb}
The computational domain consists of a unit cell comprising the layer stack shown in Fig.~\ref{fig:1}b.
We use periodic boundary conditions in the $x$-direction and assume the top and bottom ($y$-direction) to be filled with infinite half-spaces of glass and air, respectively, which is numerically treated with perfectly matched layers (PMLs) as transparent boundary conditions. For the $z$-direction we assume translational invariance of the geometry.
In a real solar cell, the glass has a finite thickness in the order of millimetres, which cannot be efficiently handled by full-field simulations.
To account for the air-glass interface on top of the solar cell, we correct for the initial reflection at this interface, which is around 4\% for normal incidence.
For the 2D simulations, the solar cell stack is discretised with an unstructured, triangular mesh with element side lengths between $3$\,nm and $50$\,nm, and we use polynomials of degree 3 to approximate the solution within each element.
The solar spectrum is sampled in the range of $\lambda_1 = 300$\,nm to $\lambda_2 = 900$\,nm with $10$\,nm step size.
The incident light is modelled as a plane wave incident from the top, i.e., propagating from $+y$ to $-y$.
The used material properties are specified in the ESI in Section \ref{supp:sec:opt_params}.
They consist of tabulated $n,k$ values obtained from various sources.
The simulation yields the local absorption density $\mathcal{A}_\text{gen}$ which is numerically integrated according to Eqs.\ \eqref{eq:photogeneration} and \eqref{eq:absorptance} to obtain the photogeneration rate $G(\mathbf{x})$ and the absorptance $A_\text{gen}$. Likewise the current densities $J_{\text{gen}}$, $J_{\text{par}}$ and $J_{R}$ are obtained by numerically integrating $A_{\text{gen}}$, $A_{\text{par}}$ and $R$ according to \eqref{eq:J_gen}.
The numerical settings for the optical simulations are chosen such that a relative numerical accuracy of better than
$10^{-3}$ is obtained for the exported photogeneration profile and the calculated maximal achievable current density $J_{\text{gen}}$.
Section \ref{supp:convergence-scan} of the ESI contains a convergence scan for both of these outputs.

\subsubsection*{Electronic model and simulation}

We employ a vacancy-assisted drift-diffusion model for the electronic simulations to describe the charge transport in the solar cells which is detailed in the ESI.
The movement of electrons and holes is considered in the ETL, PVK layer, and HTL.
Within the perovskite layer, the dynamics of ionic vacancies are also taken into account.\autocite{Abdel2021Model, Abdel2024Thesis}
Charge carrier motion is governed by drift-diffusion equations, which are self-consistently coupled to the Poisson equation via the electrostatic potential.\autocite{Farrell2017}
These equations are solved using a time-implicit finite volume scheme \autocite{Abdel2023Existence} with the excess chemical potential flux scheme for the current density approximation,\autocite{Abdel2021Flux} implemented in \texttt{ChargeTransport.jl},\autocite{ChargeTransport} which builds on the finite volume solver \texttt{VoronoiFVM.jl}.\autocite{VoronoiFVM}
The finite volume method has the major advantage of correctly reflecting physical phenomena such as local conservativity of fluxes and consistency with thermodynamic laws.\autocite{ChainaisHillairet2019, Farrell2017}
For the time discretization, we rely on an implicit Euler method.
The resulting non-linear system is solved using a damped Newton method, with the associated linear systems solved via the sparse direct solver \texttt{UMFPACK}.\autocite{Davis2004}
We generate a boundary conforming Delaunay triangulation of the computational domain using \texttt{Triangle},\autocite{Shewchuk1996} which allows to define the dual Voronoi mesh, providing the control volumes for the finite volume method.
Particular attention is paid to accurately resolving the internal material interfaces, as shown in Fig.~\ref{fig:1}c (right).
The spatial mesh contains between $47 \; 122$ nodes (planar) and $143 \; 713$ nodes (textured with $h_{\text{T}} = 750 $\,nm).
The temporal mesh for the voltage scan protocol is build adaptively: the time step size is dynamically adjusted based on convergence behaviour, with minimum and maximum step sizes of  $\Delta t_{\text{min}} = 6.0 \times 10^{-8}$\,s and $\Delta t_{\text{max}} = 8.0 \times 10^{-8}$\,s (for fast scans), resulting in approximately $150$ time steps for the forward scan.

\subsubsection*{Combining the optical and electronic model}

As illustrated in Fig.~\ref{fig:1}c (left), the simulation workflow begins with \texttt{JCMsuite}, which solves the time-harmonic Maxwell equations and computes the optical photogeneration rate in a post-processing step.
This rate is then interpolated onto a uniform $300 \times 1000$ Cartesian grid (Fig.~\ref{fig:1}c, middle) and used as input for the electronic simulations performed with \texttt{ChargeTransport.jl} (Fig.~\ref{fig:1}c, right).
Specifically, the photogeneration rate acts as a source term in the electron and hole continuity equations.
For this purpose, the optical input data is linearly interpolated via \texttt{Interpolations.jl}\autocite{Interpolations2025} and then further mapped onto the finite volume nodes.
Details of both models, along with all physically relevant material parameters, are provided in the ESI.
The simulation codes to reproduce the opto-electronic results are available in the associated data publication linked to this manuscript.\autocite{Abdel2025ElectronicData}

\subsubsection*{Author contributions}
D.A. prepared, performed and analysed the opto-electronic simulations.
J.R. prepared, performed and analysed the optical simulations.
D.A., P.F., J.F., and P.J. prepared the data publication.
P.J. gave advice on the opto-electronic software.
T.K. gave advice on the discussion of the open-circuit voltage enhancement.
C.B., S.B., and K.J. supervised the optical simulations.
P.F. and J.F. supervised the opto-electronic simulations.
D.A. and P.F. coordinated the project.
D.A., C.B., P.F., K.J, and J.R. wrote the initial manuscript, and all authors participated in proofreading and correcting the manuscript.
C.B., S.B., P.F., and K.J. initiated the project.

\subsubsection*{Conflicts of interest}
The authors declare no conflicts of interest.

\subsubsection*{Data availability}
Alongside our manuscript, we provide a data repository.\autocite{Abdel2025ElectronicData}
This repository transparently and reproducibly documents all simulations performed with the software package \texttt{ChargeTransport.jl}\autocite{ChargeTransport} for generating the opto-electronic results.
It includes all scripts and data files necessary to reproduce the opto-electronic figures.

\subsubsection*{Acknowledgements}
This project was supported by the Leibniz competition 2020 (NUMSEMIC, J89/2019) as well as the Deutsche Forschungsgemeinschaft (DFG, German Research Foundation) under Germany's Excellence Strategy -- The Berlin Mathematics Research Center MATH+ (EXC-2046/1, project ID: 390685689).
It also has received funding from the German Federal Ministry of Education and Research (BMBF Forschungscampus  MODAL, project number 05M20ZBM).
We thank Johannes Sutter for providing the SEM images of the studied layer stacks, which were taken during his time at HZB.
Further, we thank Martin Hammerschmidt, Lin Zschiedrich, and Phillip Manley from JCMwave GmbH for fruitful discussions and support.
The optical simulations were obtained in the framework of the Berlin Joint Lab for Optical Simulations for Energy Research (BerOSE) of Helmholtz-Zentrum Berlin für Materialien und Energie, Zuse-Institut Berlin, and Freie Universitat Berlin.

\printbibliography

\makeatletter\@input{xxESI.tex}\makeatother

\end{document}


\title{
\hspace*{5.0cm} \textcolor{blue}{Electronic Supplementary Information}\newline
\hspace*{0.6cm} How nanotextured interfaces influence the electronics in perovskite solar cells
}

\author{Dilara Abdel}
\affiliation{Numerical Methods for Innovative Semiconductor Devices, Weierstrass Institute for Applied Analysis and Stochastics (WIAS), Berlin, Germany}
\author{Jacob Relle}
\affiliation{Dept.\ Optics for Solar Energy, Helmholtz-Zentrum Berlin für Materialien und Energie GmbH, Berlin, Germany}
\affiliation{Computational Nano Optics, Zuse Institute Berlin, Berlin, Germany}
\author{Thomas Kirchartz}
\affiliation{IMD-3 Photovoltaics, Forschungszentrum Jülich GmbH, Jülich, Germany}
\affiliation{University of Duisburg-Essen, Duisburg, Germany}
\author{Patrick Jaap}
\affiliation{Numerical Mathematics and Scientific Computing, Weierstrass Institute for Applied Analysis and Stochastics (WIAS), Berlin, Germany}
\author{Jürgen Fuhrmann}
\affiliation{Numerical Mathematics and Scientific Computing, Weierstrass Institute for Applied Analysis and Stochastics (WIAS), Berlin, Germany}
\author{Sven Burger}
\affiliation{Computational Nano Optics, Zuse Institute Berlin, Berlin, Germany}
\affiliation{JCMwave GmbH, Berlin, Germany}
\author{Christiane Becker}
\email[E-mail: ]{christiane.becker@helmholtz-berlin.de}
\affiliation{Dept.\ Optics for Solar Energy, Helmholtz-Zentrum Berlin für Materialien und Energie GmbH, Berlin, Germany}
\affiliation{Hochschule für Technik und Wirtschaft Berlin, Berlin, Germany}
\author{Klaus Jäger}
\email[E-mail: ]{klaus.jaeger@helmholtz-berlin.de}
\affiliation{Dept.\ Optics for Solar Energy, Helmholtz-Zentrum Berlin für Materialien und Energie GmbH, Berlin, Germany}
\affiliation{Computational Nano Optics, Zuse Institute Berlin, Berlin, Germany}
\author{Patricio Farrell}
\email[E-mail: ]{patricio.farrell@wias-berlin.de}
\affiliation{Numerical Methods for Innovative Semiconductor Devices, Weierstrass Institute for Applied Analysis and Stochastics (WIAS), Berlin, Germany}

\maketitle

\tableofcontents


\vspace*{2.0cm}

This \emph{Electronic Supplementary Information} (ESI) provides additional insights into the opto-electronic properties of nanotextured perovskite solar cells.
In Sections \ref{supp:optical-model} and \ref{supp:Electronic-Model}, we provide further details on the optical and electronic models, as well as the simulation methodology.
All relevant material parameters used in this study are listed.
Section \ref{supp:addit-el-plots} presents additional simulation results, including charge carrier densities and electric fields.
Finally, Section \ref{supp:convergence-scan} provides a convergence scan for the photogeneration profile and the calculated maximal achievable short circuit current density.

\section{Optical model and simulation} \label{supp:optical-model}
We model the interaction of the electromagnetic radiation from the Sun with the solar cell by using a time-harmonic formulation of the Maxwell equations, applied to the setup shown in Fig.\ \ref{fig:1}b.
The problem is discretised via the finite element method (FEM), as implemented in the software \texttt{JCMsuite}.\cite{pomplun:2007}
This method uses a finite dimensional function space to solve a weak version of Maxwell's equations on a discretised mesh.

\subsection{Model description}

The optical model is based on the time-harmonic Maxwell equations, with the electric field $\mathbf{E}$ as the primary unknown:
\begin{equation}
\nabla \times \mu^{-1}\nabla \times \mathbf{E} - \omega^2\epsilon \mathbf{E} = i\omega \mathbf{j}^{\text{imp}},
\end{equation}
where $\mu$ is the magnetic permeability $\mu$, $\epsilon=\epsilon_0\epsilon_r$ is the complex electric permittivity, with $\epsilon_0$ denoting the vacuum permittivity and $\epsilon_r$ the relative permittivity.
The angular frequency $\omega$ is related to the vacuum wavelength $\lambda$ by $\omega=2\pi c / \lambda$, and $\mathbf{j}^{\text{imp}}$ denotes the impressed current density.\cite{jin:2014}

We assume transparent boundary conditions on top and bottom of the device, which we implement via perfectly matched layers (PML).
Bloch-periodic boundary conditions are applied in horizontal directions.\cite{pomplun:2007}
The source term is modelled by a plane wave solution in the exterior domain
\begin{equation}
    \mathbf{E} = \mathbf{E}_0\exp(i\mathbf{k}\mathbf{x}),
\end{equation}
where the wave vector $\mathbf{k}$ and the field amplitude $\mathbf{E}_0$ are determined by the angle of incidence $\phi$ and the polarization of the wave.
In our simulations, the wave enters at normal incidence from the top of the device, i.e., $\phi = 0$.
Therefore, the usual $sp$-polarization is expressed via polarization in either the $x$- or $z$-direction.
To account for the unpolarised nature of sunlight, each wavelength is modelled twice: once with $x$-polarisation and once with $z$-polarization.
The results are then averaged, assuming that both polarizations contribute equally to the total power.
For simplicity, the total incoming power is normalised for each wavelength.
The resulting optical response is then rescaled according to the corresponding irradiance $\Phi_e$ of the AM1.5G solar reference spectrum.
The spectral irradiance $\Phi_e$ follows the ASTM G-173-03 standard,\cite{NREL} provided by the U.S. Department of Energy (DOE)/NREL/ALLIANCE.
Additional details on the modelling approach can be found in the solver documentation.\cite{pomplun:2007}

\subsection{Optical material properties} \label{supp:sec:opt_params}

Wavelength-dependent complex refractive index data $n+ik$ are given for all materials in the relevant wavelength range of the solar spectrum.
To interpolate the data and calculate $\epsilon_r=\left(n+ik\right)^2$ for the desired wavelengths, we use the Python package \texttt{dispersion}.\cite{Dispersion}
The underlying $n,k$ datasets were taken from various literature sources, which are summarized in Tab.\ \ref{tab:nkdata}.

\begin{table}[!ht]
    \caption{\label{tab:nkdata}
    Used materials, their thickness in the simulation and references of the used $n,k$ datasets.
    *For the perovskite layer (PVK), the thickness denotes the \emph{effective thickness}, i.e., the thickness of a planar layer with the same volume.
    }
    \centering
    \footnotesize
    \begin{tabular}{p{2.2cm} p{2.2cm} p{5.5cm}}
    \hline
    Material & Thickness & Source of $n,k$ dataset\\
    \hline
        Glass & $1000$\,nm &Delivered with \texttt{GenPro4} \cite{GenPro4}\\
        ITO   & \phantom{0}$135$\,nm &L.\ Mazzarella \emph{et al.}\cite{Mazzarella:18}\\
        PTAA  & \phantom{00}$10$\,nm &Delivered with \texttt{GenPro4}\cite{GenPro4}\\
        PVK   & \phantom{0}$400$\,nm* &J.A.\ Guerra \emph{et al.}\cite{nk_pero}\\
        C$_{60}$& \phantom{00}$30$\,nm &D.\ Menzel \emph{et al.}\cite{nk_C_60}\\
        Cu    & \phantom{0}$100$\,nm  &P.B.\ Johnson and R.W.\ Christy\cite{nk_C_60}\\
        \hline
    \end{tabular}
\end{table}

\section{Electronic model and simulation} \label{supp:Electronic-Model}

We model the charge transport using a vacancy-assisted drift-diffusion model within a three-layer solar setup as shown in Fig.\ \ref{fig:3}a.
For clarity, we denote the total three-layer device geometry by $\Omega$, the electron transport layer (ETL) by $\Omega_{\text{ETL}}$, the intrinsic perovskite absorber (PVK) layer  by $\Omega_{\text{PVK}}$, and the hole transport layer (HTL) by $\Omega_{\text{HTL}}$.
Electrons and holes are allowed to migrate throughout the entire device. In contrast, vacancy migration is restricted to the perovskite layer $\Omega_{\text{PVK}}$, where we account for volume exclusion effects.
The system is discretized using an implicit-in-time finite volume scheme, which ensures local flux conservation and consistency with thermodynamic laws -- key advantages of the finite volume method (FVM).
Moreover, the existence and boundedness of both weak and discrete FVM solutions have been rigorously established.\cite{Abdel2023Existence, Abdel2024Analysis}
For a detailed discussion of the model and the discretization, we refer to previous studies.\cite{Abdel2023VolumeExclusion, Abdel2021Model, Abdel2024Thesis}

\subsection{Model description}

\textbf{Charge transport equations}
The charge transport model considers the electric potential $\psi$ and the
quasi Fermi potentials of moving charge carriers $\varphi_\alpha$ as unknowns.
Here, the indices $\alpha \in \{ \electrons, \holes, \ions \}$ refer to the electrons, holes, and anion vacancies.
Within the perovskite layer $\Omega_{\text{PVK}}$, carrier transport is governed by the following equations
\begin{subequations}
\begin{align}
    \quad\quad - \nabla \cdot (\varepsilon_s \nabla \psi ) &=
    q \Bigl( z_\electrons n_\electrons + z_\holes n_\holes + z_\ions n_\ions  + C( \mathbf{x} )\Bigr), \label{eq:poisson}\\
    \quad\quad z_\electrons q \partial_t n_\electrons + \nabla\cdot \mathbf{j}_\electrons &=   z_\electrons q \left(G(\mathbf{x}) - R( n_\electrons, n_\holes)\right), \label{eq:electrons}\\
    \quad\quad z_\holes q \partial_t n_\holes + \nabla\cdot \mathbf{j}_\holes &=  z_\holes q \left(G(\mathbf{x}) - R( n_\electrons, n_\holes)\right), \label{eq:holes}\\
    z_\ions q \partial_t n_\ions + \nabla\cdot \mathbf{j}_\ions &\;= 0.
\end{align}
\end{subequations}
Only electrons and holes are considered in the transport layers  $\Omega_{\text{ETL}} \cup  \Omega_{\text{HTL}}$, meaning the model reduces to equations \eqref{eq:poisson}-\eqref{eq:holes} with a modified space charge density in that case.
The dielectric permittivity is defined as $\varepsilon_s = \varepsilon_0 \varepsilon_r $, where $\varepsilon_0$ is the vacuum permittivity and $\varepsilon_r$ the relative material permittivity.
The charge numbers are given by $z_\electrons = -1$, $z_\holes = 1$, and $z_\ions = 1$.
The doping and the mean vacancy concentration are included in the quantity $C$. Moreover, $G$ denotes the optical photogeneration rate and $R$ the recombination processes.
Finally, the current density $\mathbf{j}_\alpha$ describes the motion of charge carriers
\begin{align}
    \mathbf{j}_\alpha  = -z_\alpha q \left( D_\alpha\left(\frac{n_\alpha}{N_\alpha} \right)  \nabla n_\alpha + z_\alpha \mu_\alpha n_\alpha \nabla \psi \right), \quad  \alpha \in \{\electrons, \holes, \ions \},
\end{align}
where $D_\alpha$ represents the non-linear diffusion coefficient and $\mu_\alpha$ the mobility of carriers.
We can link the charge carrier densities $n_\alpha$ to the potentials, defining the set of unknowns, for $\alpha \in \{\electrons, \holes, \ions \}$ via
\begin{align} \label{eq:stateeq}
    n_\alpha = N_\alpha \mathcal{F}_\alpha \Bigl(\eta_\alpha(\varphi_\alpha, \psi)\Bigr),
    \quad
    \eta_\alpha
    = \frac{\varPhi_\alpha}{k_BT}
    = \frac{z_\alpha \left(E_\alpha - E_{\text{F}, \alpha}\right)}{k_BT}
    = z_\alpha \frac{q (\varphi_\alpha - \psi) + E_{\alpha, 0}}{k_B T}.
\end{align}
Here, $T$ is the temperature, $k_B$ is the Boltzmann constant, and $N_{\text{c}} = N_\electrons$ and $N_{\text{v}} = N_\holes$ are the effective densities of state of the conduction band and valence band.
The parameters $E_{\text{c}, 0} = E_{\electrons, 0}$ and $E_{{\text{v}}, 0} = E_{\holes, 0}$ denote the intrinsic band edge energies of the conduction and valence bands for electrons and holes.
In contrast, $E_{\ions, 0}$ corresponds to an intrinsic energy level of vacancies.
The quantities $\varPhi_\electrons = z_\electrons (E_{\text{c}} - E_{\text{F}, \electrons})$ and $\varPhi_\holes = z_\holes (E_{\text{v}} - E_{\text{F}, \holes})$ are the energy off-sets of electrons and holes, respectively.
Moreover,  $E_{\text{F, \electrons}} = - q \varphi_\electrons$ is the electron quasi Fermi level and $E_{\text{F, \holes}} = - q \varphi_\holes$ is the hole quasi Fermi level.
The statistics function $\mathcal{F}_\alpha$ varies depending on the type of charge carrier.
For electrons and holes, $\mathcal{F}_\electrons$ and $\mathcal{F}_\holes$ depend on the semiconductor material: the Fermi-Dirac integral of order 1/2 is used for inorganic materials, while organic materials require the Gauss-Fermi integral.
In the limit of $n_\alpha \ll N_\alpha$ both integrals are well approximated by $\mathcal{F}_\electrons = \mathcal{F}_\holes \approx \exp$ which is the case here.
In contrast, for vacancies the function $\mathcal{F}_\alpha(\eta) = 1/(\exp(-\eta) + 1)$, called Fermi-Dirac integral of order $-1$, ensures the correct limitation of ion depletion.
In \eqref{eq:stateeq}, $N_\ions$ denotes a saturation density, limited by the density of available lattice sites in the crystal.

\textbf{Recombination and photogeneration}
The recombination rate $R$ on the right-hand side of the electron and hole mass balance equations \eqref{eq:electrons}-\eqref{eq:holes} is modelled by the sum of the dominant recombination processes: Shockley-Read-Hall (SRH), radiative, and surface recombination with
\begin{align}
  R(n_\electrons, n_\holes) = \sum_{r} R_r(n_\electrons, n_\holes), \quad r \in \{ \text{SRH}, \text{rad}, \text{surf} \}.
\end{align}
Each recombination contribution $R_r$ follows the general form
\begin{align}
  R_{r}(n_\electrons, n_\holes) = r_r(n_\electrons, n_\holes) n_\electrons n_\holes \left(1- \exp \left(\frac{q\varphi_\electrons - q\varphi_\holes}{k_B T}\right)\right).
\end{align}
The process-dependent, non-negative rate functions $r_r$ are defined as follows.
First, for SRH recombination, we have
\begin{equation}
  r_{\text{SRH}}(n_\electrons, n_\holes) = \frac{1}{\tau_\holes (n_\electrons + n_{\electrons, \tau}) + \tau_\electrons (n_\holes + n_{\holes, \tau})},
\end{equation}
where $\tau_\electrons, \tau_\holes$ are the carrier lifetimes and $n_{\electrons, \tau}, n_{\holes, \tau}$ are reference carrier densities.
The SRH model assumes a single mid-gap trap level, consistent with prior work and the literature.\cite{LeCorre2022,Diekmann2021,Clarke2022}
While more detailed models that incorporate multiple defect levels exist, these are beyond the scope of the present study.
Second, for the radiative recombination the prefactor is given by
\begin{equation} \label{eq:R-Rad}
  r_{\text{rad}} (n_\electrons, n_\holes) = r_{0, \text{rad}}
\end{equation}
for a constant rate coefficient $r_{0, \text{rad}}$.
Lastly, the surface recombination can be modelled as
\begin{equation} \label{eq:surface-r}
	r_{\text{surf}}(n_\electrons, n_\holes) = \frac{1}{\frac{1}{\nu_\holes} (n_\electrons + n_{\electrons, \tau}) + \frac{1}{\nu_\electrons} (n_\holes + n_{\holes, \tau})},
\end{equation}
where $\nu_\electrons$, $\nu_\holes$ are the surface recombination velocities.
Unless stated otherwise, the photogeneration $G$, defined in the accompanying paper, is obtained as the solution of the time-harmonic Maxwell equations described in Section \ref{supp:optical-model}.

\textbf{Boundary and initial conditions}
Following the device geometry depicted in Fig.\ \ref{fig:3}a, we assume that at the top and the bottom of the architecture, we have the metal-semiconductor interfaces, denoted by $\Gamma^{\text{D}}$.
The left and right boundaries are denoted by $\Gamma^{\text{N}}$.
We model the boundary conditions for $t \geq 0$ as follows:
\begin{alignat}{2}
  \psi(\mathbf{x}, t) =\psi_0(\mathbf{x}) + V(\mathbf{x}, t), \;\; \varphi_\electrons(\mathbf{x}, t)  = \varphi_\holes(\mathbf{x}, t) &= V (\mathbf{x}, t), \quad
  ~
  &&\mathbf{x} \in \Gamma^{\text{D}},  \label{eq:Dirichlet-cond-physical} \\[0.5ex]
  ~
  ~
  \nabla \psi(\mathbf{x}, t) \cdot \boldsymbol{\nu}(\mathbf{x}) = \mathbf{j}_\electrons(\mathbf{x}, t) \cdot \boldsymbol{\nu}(\mathbf{x}) = \mathbf{j}_\holes(\mathbf{x}, t) \cdot \boldsymbol{\nu}(\mathbf{x}) &= 0, \quad
  ~
  &&\mathbf{x} \in \Gamma^{\text{N}}.
\end{alignat}
Here, $V$ corresponds to the externally applied measurement protocol, and $\boldsymbol{\nu}$ denotes the outward pointing unit normal vector to $\Gamma^{\text{N}}$.
The potential $\psi_0$, often referred to as built-in potential, will be specified in Section \ref{sec:El-Params}.
Regarding the anion vacancies, we impose no flux Neumann boundary conditions on the entire boundary of the intrinsic layer, namely
\begin{equation}\label{eq:outer-bc_anion}
  \mathbf{j}_{\ions}(\mathbf{x}, t) \cdot \boldsymbol{\nu}_\text{PVK}(\mathbf{x}) = 0, \quad
  \mathbf{x} \in \partial\Omega_{\text{PVK}},  \ t \geq 0,
\end{equation}
where $\boldsymbol{\nu}_\text{PVK}$ is the outward pointing unit normal vector to $ \partial\Omega_{\text{PVK}}$.

Lastly, we supply the system with initial conditions for $t = 0$
\begin{subequations}
    \begin{alignat}{2} \label{eq:initial-cond}
  \varphi_\electrons(\mathbf{x}, 0) = \varphi_\electrons^0(\mathbf{x}), \quad &\quad\varphi_\holes(\mathbf{x}, 0) = \varphi_\holes^0(\mathbf{x}), \quad &&\mathbf{x} \in \Omega,\\
  \varphi_\ions(\mathbf{x}, 0) &= \varphi_\ions^0(\mathbf{x}), ~ \quad &&\mathbf{x} \in \Omega_{\text{PVK}}.
\end{alignat}

\end{subequations}

The total current is obtained by integrating the sum of current densities over the device contacts, as determined from the numerical solution of the coupled system of partial differential equations.\cite{Farrell2017}
\subsection{Electronic material parameters} \label{sec:El-Params}

Table \ref{tab:sim-parameter-PSC} summarizes the electronic parameters used in all electronic simulations.
In consistency with Thiesbrummel \emph{et al.},\cite{thiesbrummel2024ion} we set the boundary values for the built-in potential as follows: at the bottom contact, $\psi_0 = -(0.05 \, \text{eV} - E_{\text{c}, 0})/q$, and at the top contact, $\psi_0 = -(-0.05 \, \text{eV} - E_{\text{v}, 0})/q$, thereby including the energy offset $0.05$\,eV between the metal and the transport layer.
Note that the scan protocol is applied at the top contact.
Furthermore, we set the doping term on the right-hand side of the Poisson equation \eqref{eq:poisson} to $C = C_\electrons$ in the electron transport layer, $C = - C_\ions$ in the intrinsic perovskite layer and $C = - C_\holes$ in the hole transport layer.

\begin{table}[!ht]
    \caption{Parameter values for the simulation of a single-junction ``83-17 triple cation'' perovskite solar cell at a temperature $T = 300$\,K with $C_{60}$ as electron transport layer material and PTAA as hole transport layer material, mainly based on Thiesbrummel \emph{et al.}, \cite{thiesbrummel2024ion} for which a data publication is available.
     *For the perovskite layer, the thickness denotes the \emph{effective thickness}, i.e., the thickness of a planar layer with the same volume.}
    \label{tab:sim-parameter-PSC}
    \centering
   \footnotesize
  \begin{tabularx}{\columnwidth}{l c c c c c c}
    \hline
    \hspace*{0.1cm}\textbf{Physical quantity} 	& \hspace*{0.13cm} \textbf{Symbol}   \hspace*{0.18cm}	 & \hspace*{1.3cm}		 & \hspace*{1.2cm} \textbf{Value}  \hspace*{1.3cm} & \hspace*{1.3cm}	  &  \hspace*{0.7cm}	\textbf{Unit} \hspace*{0.8cm}	& \textbf{Ref.} \\[0.0ex]
        \hline
        &&$C_{60}$ & Perovskite & PTAA \\[0.0ex]
        \hline\\[-4.0ex]
        Layer thickness &                                     &  $30$ & $400$* & $10$ & \si{\nano\metre} & \citenum{LeCorre2022,thiesbrummel2024ion} \\[0.5ex]
        Relative permittivity & $\varepsilon_r$               & $5$ & $22.0$ & $3.5$  & &  \citenum{thiesbrummel2024ion,LeCorre2022}\\[0.5ex]
        Conduction band-edge energy	& $E_{\text{c}, 0}$       & $-3.9$ & $-3.9$ & $-2.5$ & \si{\electronvolt} & \citenum{thiesbrummel2024ion,LeCorre2022}\\[0.5ex]
        Valence band-edge energy	& $E_{\text{v}, 0}$         & $-5.9$ & $-5.53$ & $-5.5$ &  \si{\electronvolt} &  \citenum{thiesbrummel2024ion,LeCorre2022}\\[0.5ex]
        Eff.\ conduction band DoS & $N_{\text{c}}$            & $ \num{1e26}$ & $ \num{2.2e24}$ & $ \num{1e26}$ &\si{\metre^{-3}} &  \citenum{thiesbrummel2024ion,LeCorre2022}\\[0.5ex]
        Eff.\ valence band DoS & $N_{\text{v}}$              & $ \num{1e26}$ &  $ \num{2.2e24}$ & $ \num{1e26}$ &\si{\metre^{-3}}&  \citenum{thiesbrummel2024ion,LeCorre2022} \\[0.5ex]
        Max.\ vacancy density & $N_\ions$                     & -- &  $ \num{1.0e27}$ & -- &\si{\metre^{-3}} \\[0.5ex]
        Doping density & $C_\electrons$                       & $\num{1.0e20} $ & 0.0 & 0.0 & \si{\metre^{-3}}\\[0.5ex]
        Doping density & $C_\holes$                           & 0.0 & 0.0&$\num{1.0e20}$ & \si{\metre^{-3}}\\[0.5ex]
        Average vacancy density & $C_\ions$                   & -- &$\num{6.0e22}$& -- &\si{\metre^{-3}} &  \citenum{LeCorre2022}\\[0.5ex]
        Electron mobility	& $\mu_\electrons$                  & $\num{1.0e-6} $ & $\num{1.0e-4}$ & $\num{1.0e-8}$&  \si{\metre^2}$\cdot$\,(\si{\V\s})$^{-1}$ & \citenum{thiesbrummel2024ion} \\[0.5ex]
        Hole mobility	& $\mu_\holes$                          & $\num{1.0e-6} $ & $\num{1.0e-4}$ & $\num{1.0e-8}$&  \si{\metre^2}$\cdot$\,(\si{\V\s})$^{-1}$ & \citenum{thiesbrummel2024ion} \\[0.5ex]
        Vacancy mobility	& $\mu_\ions$                       & -- & $\num{1.9e-12}$ & -- &  \si{\metre^2}$\cdot$\,(\si{\V\s})$^{-1}$ &  \citenum{LeCorre2022} \\[0.5ex]
        Rad.\ recombination coeff. & $r_{0, \text{rad}}$      & $\num{0.0}$ & $\num{3.0e-17} $ & $0.0$  & \si{\metre^{3}}$\cdot$\,\si{\s}$^{-1}$ & \citenum{thiesbrummel2024ion,Diekmann2021} \\[0.5ex]
        SRH lifetime, electrons & $\tau_\electrons$           & $\num{1.0e100}$  & $\num{2.0e-7} $ & $\num{1.0e100}$ & \si{\s} & \citenum{thiesbrummel2024ion, stolterfoht2018visualization} \\[0.5ex]
        SRH lifetime, holes & $\tau_\holes$                   & $\num{1.0e100}$  & $\num{2.0e-7} $ & $\num{1.0e100}$ & \si{\s} & \citenum{thiesbrummel2024ion, stolterfoht2018visualization}\\[0.5ex]
        SRH ref.\ dens.\, electrons & $n_{\electrons, \tau}$  & $\num{1.59e9}$  & $\num{4.48e10} $ & $\num{6.33}$ & \si{\metre^{-3}}\\[0.5ex]
        SRH ref.\ dens.\, holes & $n_{\holes, \tau}$          & $\num{1.59e9}$  & $\num{4.48e10} $ & $\num{6.33}$ & \si{\metre^{-3}}\\[1.8ex]  \hline
    \end{tabularx}
\end{table}

The surface recombination velocities at the perovskite/HTL interface $v_{\text{HTL}}$ and at the ETL/PVK interface are varied throughout our study.
For the reference configuration,\cite{thiesbrummel2024ion} we set $v_{\text{HTL}} = v_\electrons = v_\holes = \num{2000}$\,cm$\cdot$s$^{-1}$ and $v_{\text{ETL}} = v_\electrons = v_\holes = \num{500}$\,cm$\cdot$s$^{-1}$.
To remain consistent with the literature,\cite{LeCorre2022,thiesbrummel2024ion} we adopt the default SRH reference densities from the simulation tool \texttt{Ionmonger},\cite{Clarke2022} which was used in the aforementioned studies.

Specifically, the reference densities in the surface recombination terms at the HTL/perovskite interface are set to $n_{\electrons, \tau}= \left. n_{\electrons, \tau} \right|_{\Omega_{\text{PVK}}}$ and $n_{\holes, \tau}= \left. n_{\holes, \tau} \right|_{\Omega_{\text{HTL}}}$, while at the perovskite/ETL interface we have $n_{\electrons, \tau}= \left. n_{\electrons, \tau} \right|_{\Omega_{\text{ETL}}}$ and $n_{\holes, \tau}= \left. n_{\holes, \tau} \right|_{\Omega_{\text{PVK}}}$, which are stated in Tab.\ \ref{tab:sim-parameter-PSC}.

In Thiesbrummel \emph{et al.}, \cite{thiesbrummel2024ion} the transport layers (TLs) are assumed to be undoped.
For numerical stability, the doping in the TLs is set to $\num{1.0e20}$\,m$^{-3}$, which is small compared to the effective density of states for electrons and holes and can therefore be interpreted as effectively undoped.
The maximum vacancy density $N_{\ions}$ corresponds to a saturation limit that reflects the finite number of available lattice sites in the perovskite and can be estimated from the lattice constant.\cite{Abdel2024Thesis}

We note that for the one-dimensional planar setup, our simulations with \texttt{ChargeTransport.jl} match those obtained with \texttt{Ionmonger}, the simulation tool used in the aforementioned works.\cite{LeCorre2022, Diekmann2021,thiesbrummel2024ion}

\section{Additional electronic results} \label{supp:addit-el-plots}

In this section, we present additional figures that support the arguments made in the main text.
Section \ref{supp:recomb-current} provides the recombination current densities and Sections \ref{supp:add-el-low} and \ref{supp:add-el-ref} show additional results for the low-surface-recombination case ($C_4$) and the reference configuration ($C_1$), respectively.
For the intermediate test cases $C_2$ and $C_3$ we have similar results.

\subsection{Recombination currents} \label{supp:recomb-current}

\begin{figure*}[ht!]
  \centering
  \includegraphics[width=1.0\textwidth]{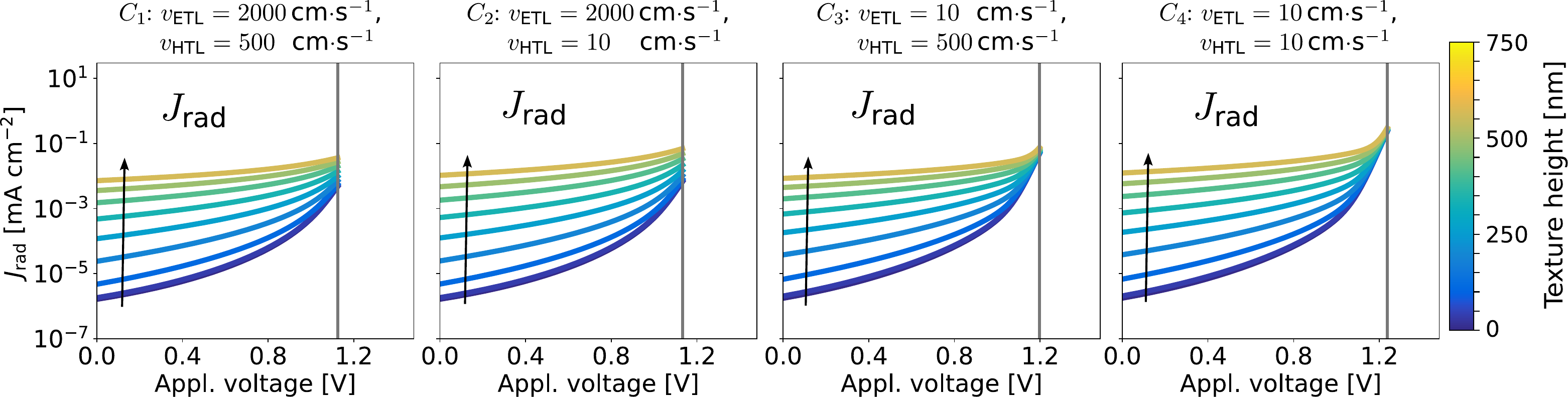}
  \caption{
    Radiative recombination current density $J_{\text{rad}}$ for all four test cases $C_1$ to $C_4$ for varying texture heights.
    The arrows indicate the direction of increasing texture height.
    }
  \label{supp:figs1}
\end{figure*}

\newpage

\begin{figure*}[ht!]
  \centering
  \includegraphics[width=1.0\textwidth]{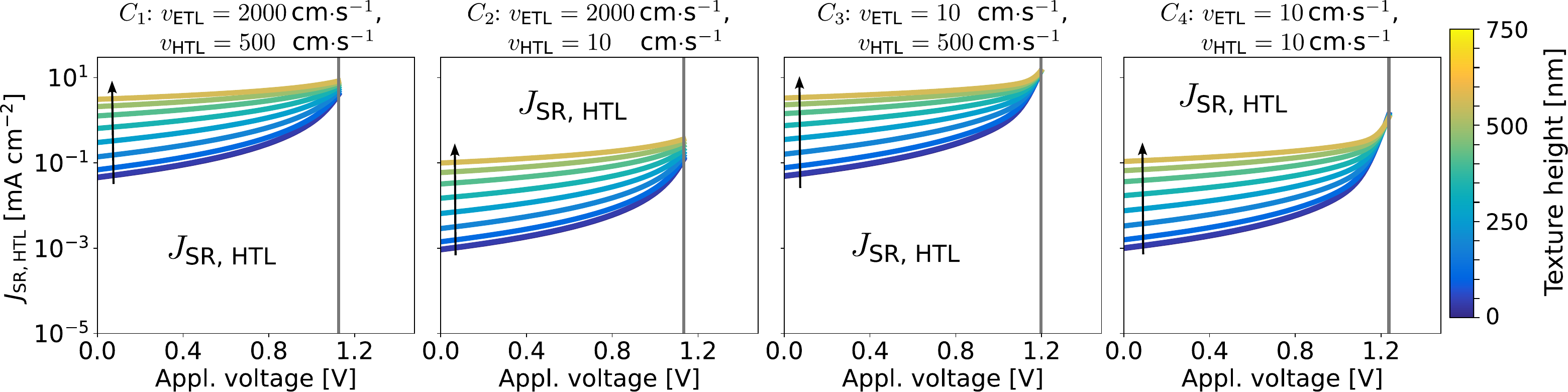}
  \caption{
    Surface recombination current density at the PVK/HTL interface $J_{\text{SR, HTL}}$ for all four test cases $C_1$ to $C_4$ for varying texture heights.
    The arrows indicate the direction of increasing texture height.
    }
  \label{supp:figs2}
\end{figure*}


\begin{figure*}[ht]
  \centering
  \includegraphics[width=1.0\textwidth]{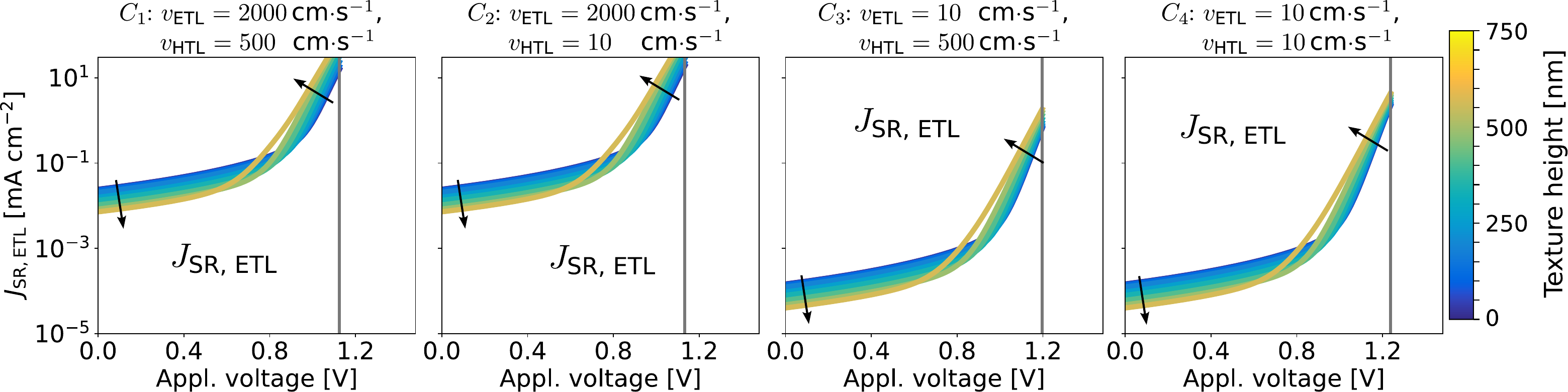}
  \caption{
    Surface recombination current density at the ETL/PVK interface $J_{\text{SR, ETL}}$ for all four test cases $C_1$ to $C_4$ for varying texture heights.
    The arrows indicate the direction of increasing texture height.
    }
  \label{supp:figs3}
\end{figure*}


\begin{figure*}[ht!]
  \centering
  \includegraphics[width=1.0\textwidth]{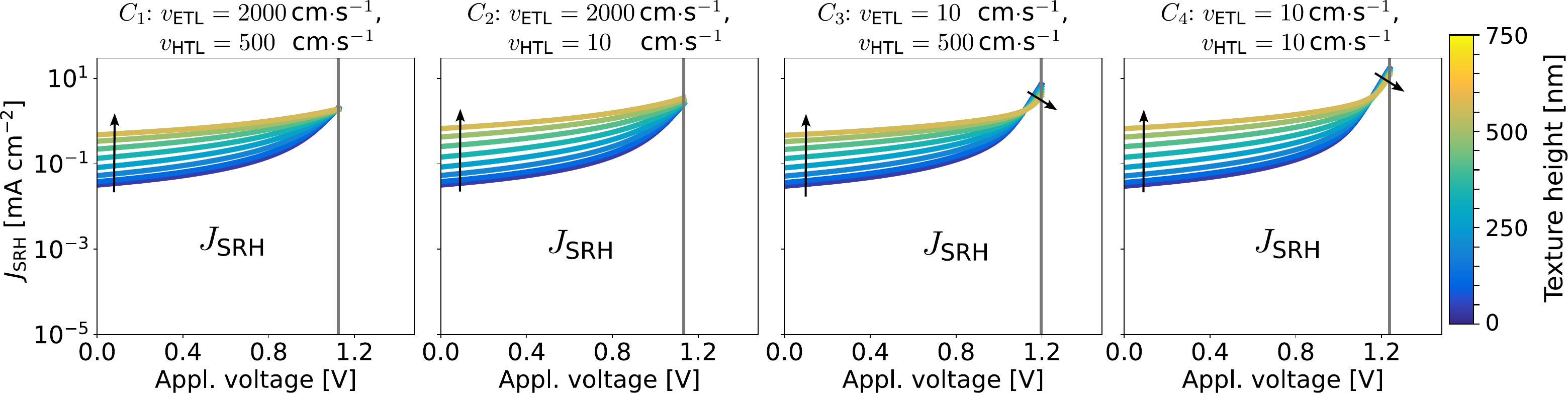}
  \caption{
    Shockley-Read-Hall recombination current density $J_{\text{SRH}}$ for all four test cases $C_1$ to $C_4$ for varying texture heights.
    The arrows indicate the direction of increasing texture height.
    }
  \label{supp:figs4}
\end{figure*}

\newpage
\vspace*{-2.5cm}
\subsection{Carrier densities and electric field for low surface recombination}  \label{supp:add-el-low}

\begin{figure*}[h!]
  \vspace*{-0.7cm}
  \centering
  \includegraphics[width=0.78\textwidth]{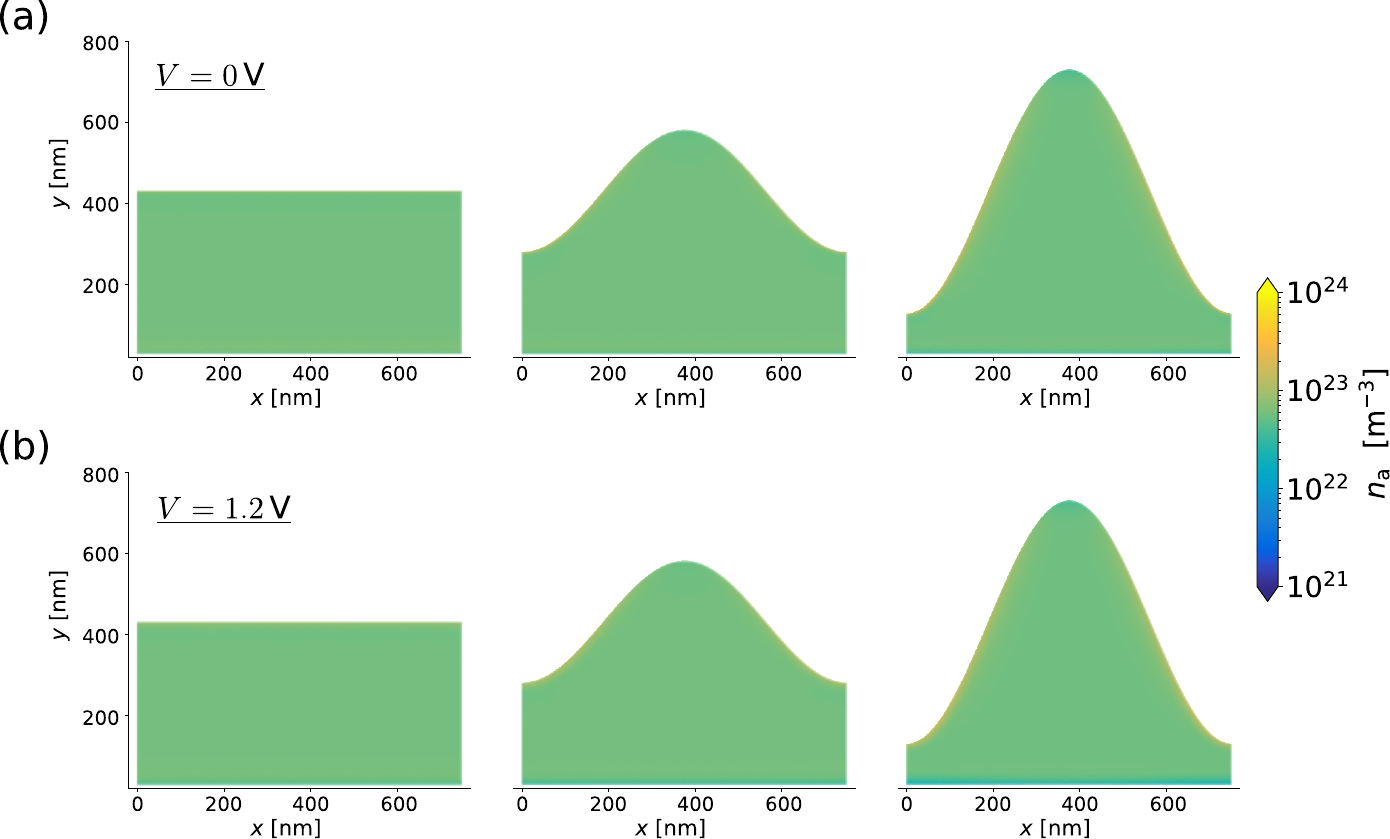}
  \caption{
    Simulated vacancy density $n_\ions$ within the PVK layer for the studied solar cell setup during the forward scan for the test case $C_4$ at (a) zero applied voltage and (b) an applied voltage near open-circuit voltage ($V = 1.2$\,V).
    We have an average vacancy density of $\overline{n_\ions} = 6.0 \times 10^{22}$\,m$^{-3}$ for all texture heights.
    }
  \label{supp:figs5}
\end{figure*}


\begin{figure*}[h!]
  \vspace*{-0.85cm}
  \centering
  \includegraphics[width=0.78\textwidth]{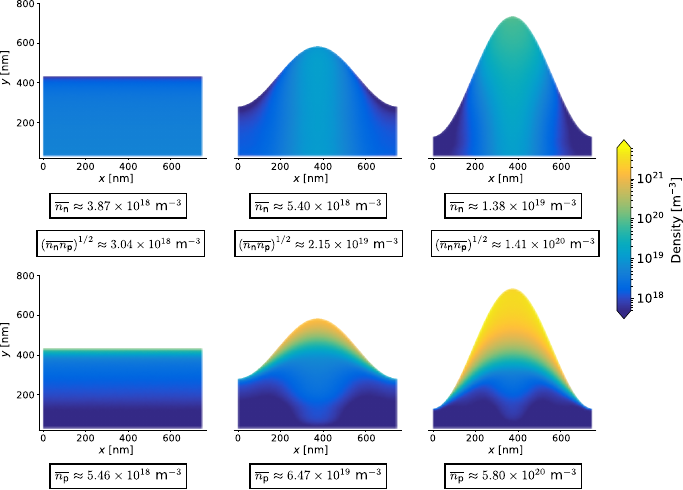}
  \caption{
      Simulated charge carrier densities of electrons $n_{\text{n}} $ and holes $n_{\text{p}}$ during the forward scan for the test case $C_4$ at an applied voltage near short-circuit conditions ($V = 0$\,V).
      The boxes indicate the integral averages of the densities over the perovskite material layer.
    }
  \label{supp:figs6}
\end{figure*}


\begin{figure*}[ht!]
  \centering
  \includegraphics[width=0.80\textwidth]{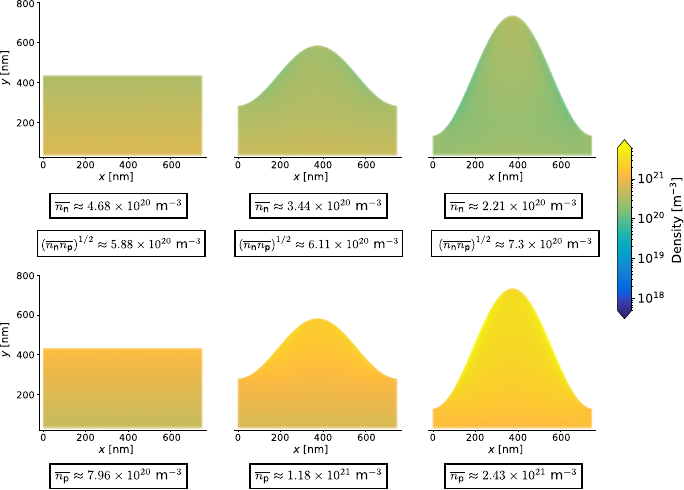}
  \caption{
      Simulated charge carrier densities of electrons $n_{\text{n}} $ and holes $n_{\text{p}}$ during the forward scan for the test case $C_4$ at an applied voltage near open-circuit ($V = 1.2$\,V).
      The boxes indicate the integral averages of the densities over the perovskite material layer.
    }
  \label{supp:figs7}
\end{figure*}


\begin{figure*}[ht!]
  \centering
  \includegraphics[width=0.8\textwidth]{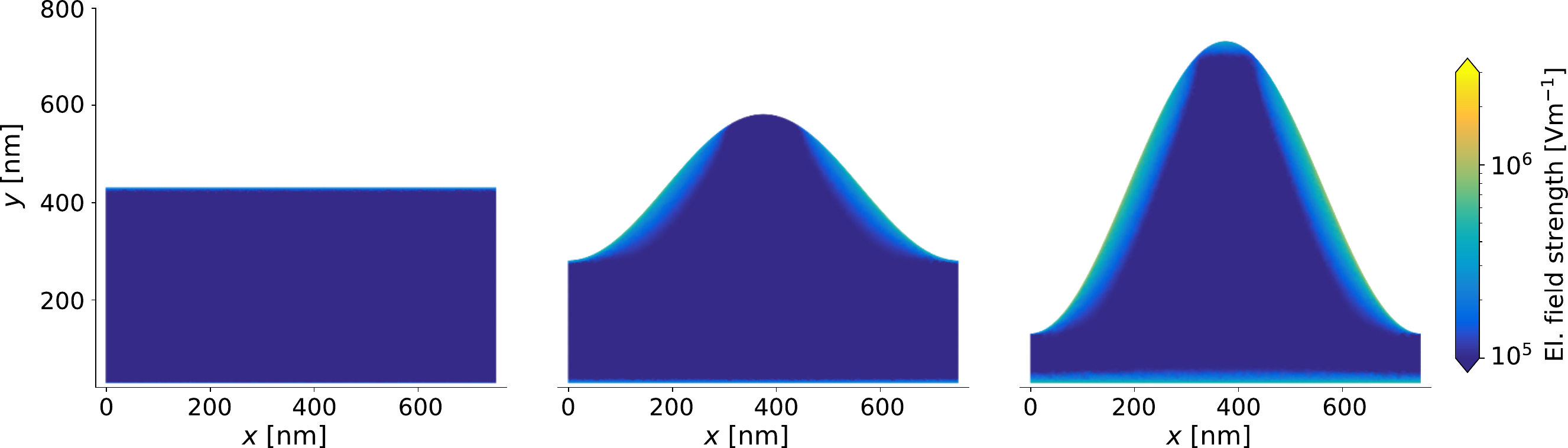}
  \caption{
    Simulated electric field for the test case $C_4$ at an applied voltage near open-circuit ($V = 1.2$\,V).
    The colour and the stream plot indicate the strength $\Vert - \nabla \psi \Vert_2 $ and the direction of the electric field, respectively.
    }
  \label{supp:figs8}
\end{figure*}

\vspace*{2.0cm}
\newpage
\subsection{Carrier densities and electric field for reference configuration} \label{supp:add-el-ref}

\begin{figure*}[h!]
  \centering
  \includegraphics[width=\textwidth]{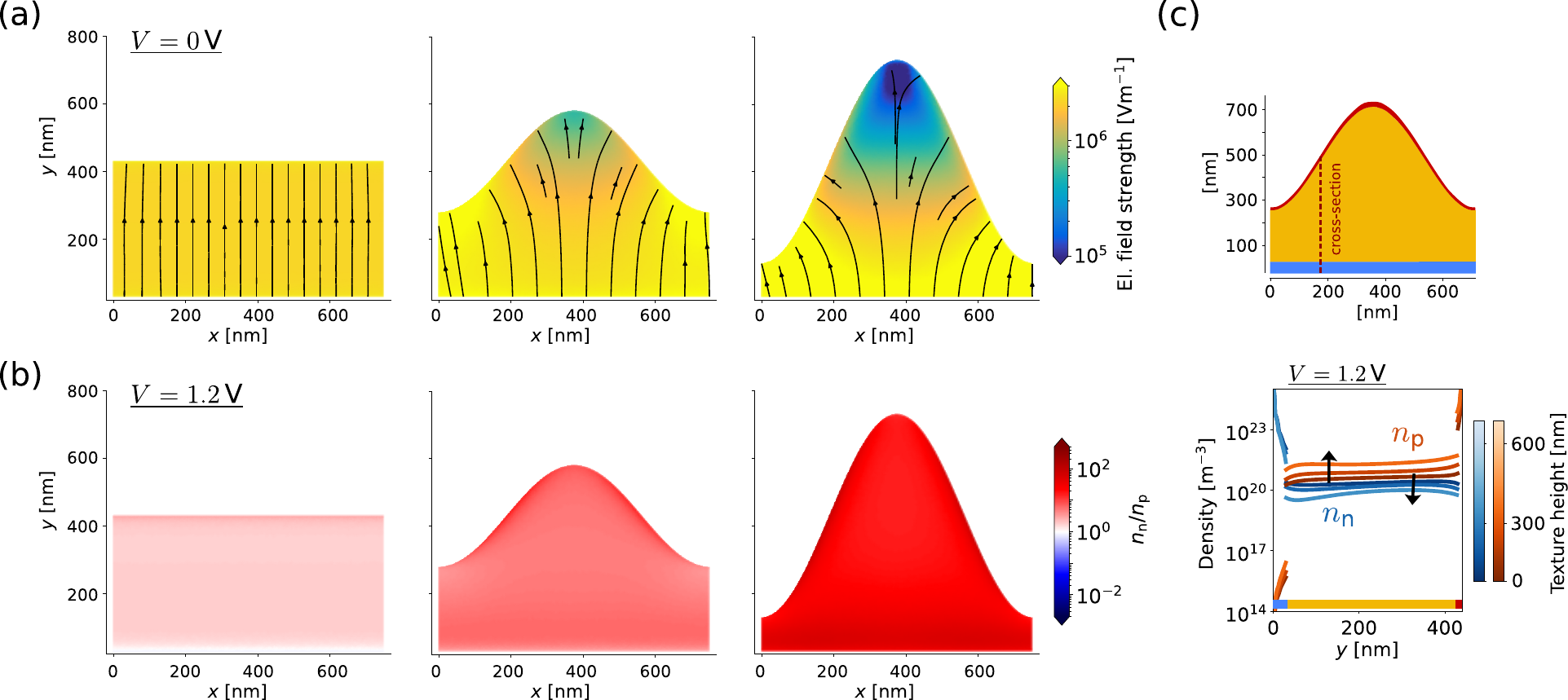}
  \caption{
    Analogously to Fig.\ \ref{fig:5}, we show the electric field and carrier densities for the test case $C_1$ (reference configuration).
    (a) Electric field for three texture heights $h_{\text{T}} = 0, 300, 600$\,nm for $V = 0$\,V applied voltage.
    The colour and the stream plot indicate the strength $\Vert - \nabla \psi \Vert_2 $ and the direction of the electric field, respectively.
    (b) The corresponding ratio between hole and electron density $n_\holes/n_\electrons$ for $V = 1.2$\,V applied voltage.
    (c) 2D device geometry with the vertical cross-section indicated (top), along which the carrier densities (bottom) are extracted. More precisely, we see one-dimensional profiles of the electron (blue) and hole densities (red) at an applied voltage $V = 1.2$\,V for varying texture height.
    In the density plot, brighter colours indicate greater texture height, with arrows showing the direction of increasing texture height.
    }
  \label{supp:figs9}
\end{figure*}

\vspace*{2.0cm}

\begin{figure*}[h!]
  \vspace*{-1.3cm}
  \centering
  \includegraphics[width=0.78\textwidth]{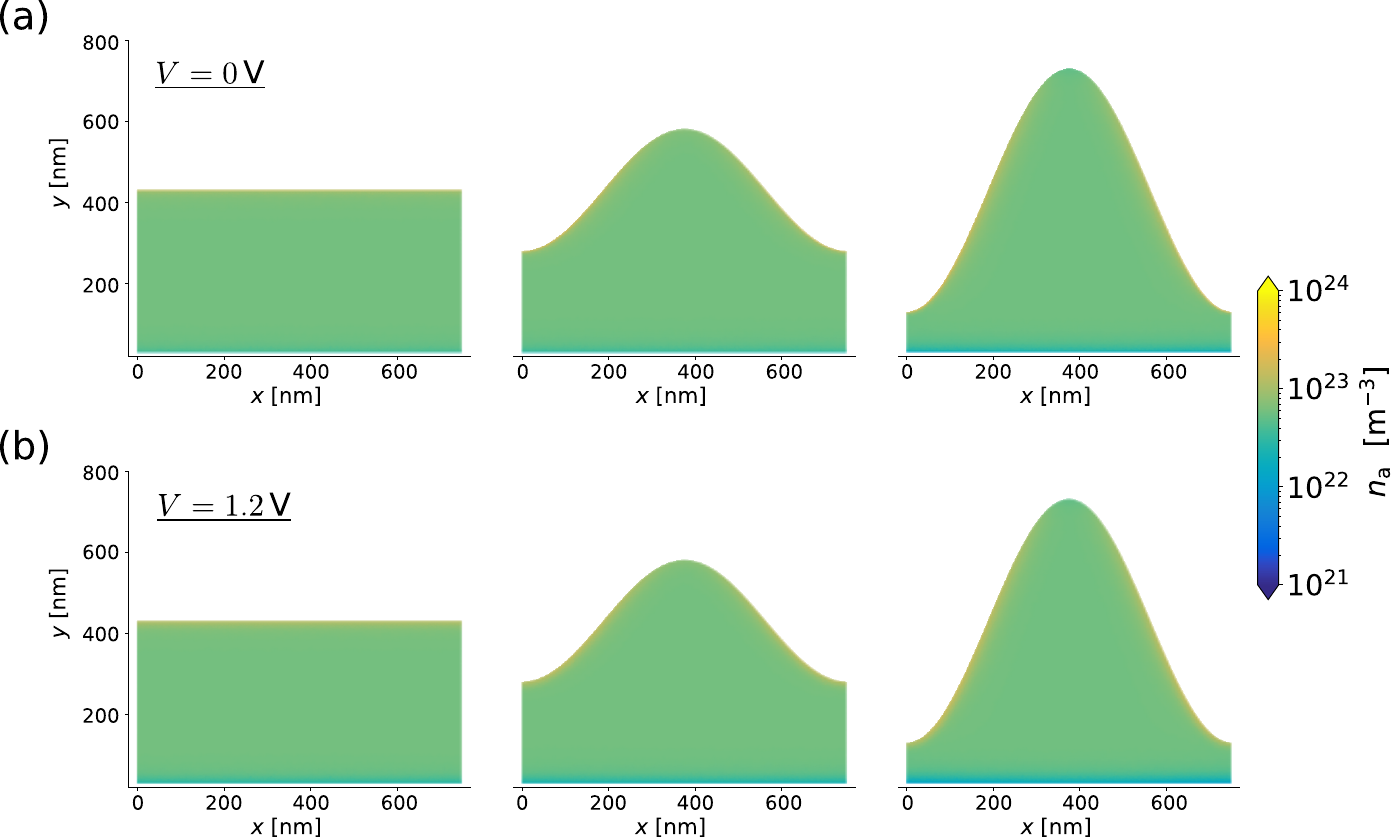}
  \caption{
      Simulated vacancy density $n_\ions$ within the PVK layer for the studied solar cell setup during the forward scan for the test case $C_1$ at (a) zero applied voltage and (b) an applied voltage near open-circuit voltage ($V = 1.2$\,V).
      We have an average vacancy density of $\overline{n_\ions} = 6.0 \times 10^{22}$\,m$^{-3}$ for all texture heights.
    }
  \label{supp:figs10}
\end{figure*}


\begin{figure*}[h!]
   \vspace*{-0.85cm}
  \centering
  \includegraphics[width=0.78\textwidth]{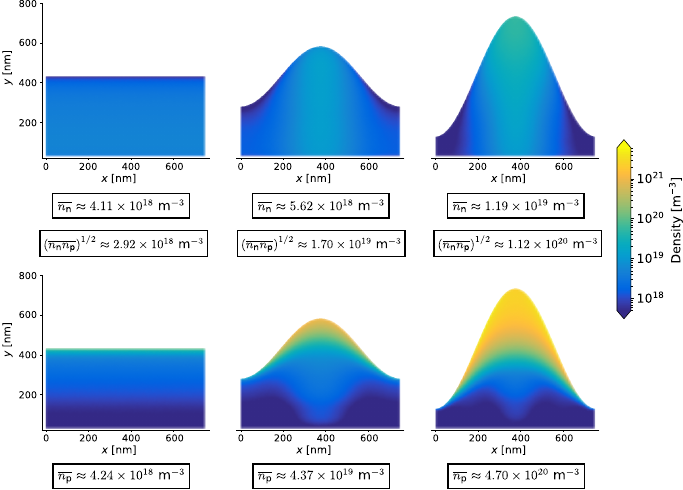}
  \caption{
      Simulated charge carrier densities of electrons $n_{\text{n}} $ and holes $n_{\text{p}}$ during the forward scan for the test case $C_1$ at an applied voltage near short-circuit conditions ($V = 0$\,V).
      The boxes indicate the integral averages of the densities over the perovskite material layer.
    }
  \label{supp:figs11}
\end{figure*}


\begin{figure*}[ht!]
  \centering
  \includegraphics[width=0.85\textwidth]{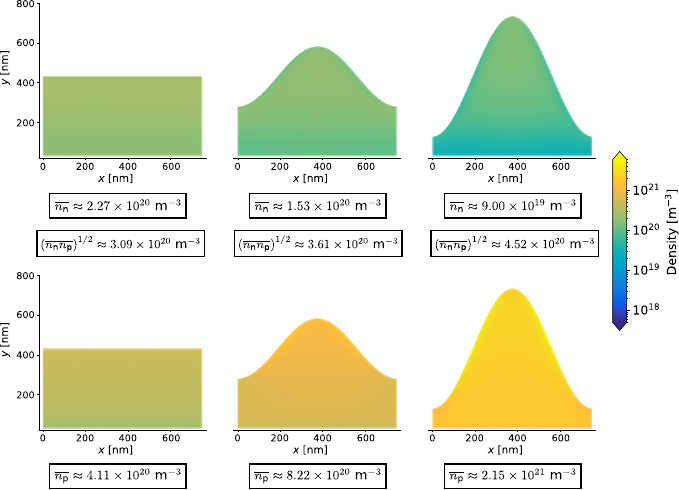}
  \caption{
      Simulated charge carrier densities of electrons $n_{\text{n}} $ and holes $n_{\text{p}}$ during the forward scan for the test case $C_1$ at an applied voltage near open-circuit voltage ($V = 1.2$\,V).
      The boxes indicate the integral averages of the densities over the perovskite material layer.
    }
  \label{supp:figs12}
\end{figure*}


\begin{figure*}[ht!]
  \centering
  \includegraphics[width=0.8\textwidth]{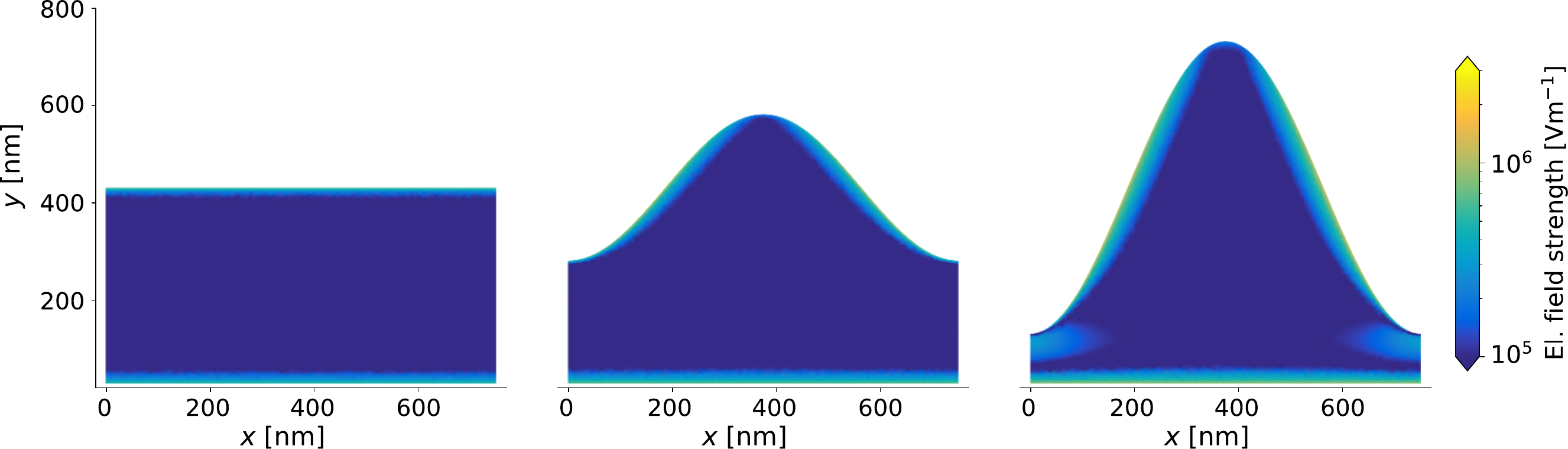}
  \caption{
    Simulated electric field for the test case $C_1$ at an applied voltage near open-circuit ($V = 1.2$\,V).
    The colour and the stream plot indicate the strength $\Vert - \nabla \psi \Vert_2 $ and the direction of the electric field, respectively.
    }
  \label{supp:figs13}
\end{figure*}

\newpage
\section{Optical convergence scan} \label{supp:convergence-scan}

We check the convergence of the optical simulations.

\begin{figure*}[ht]
  \centering
  \includegraphics[width=\textwidth]{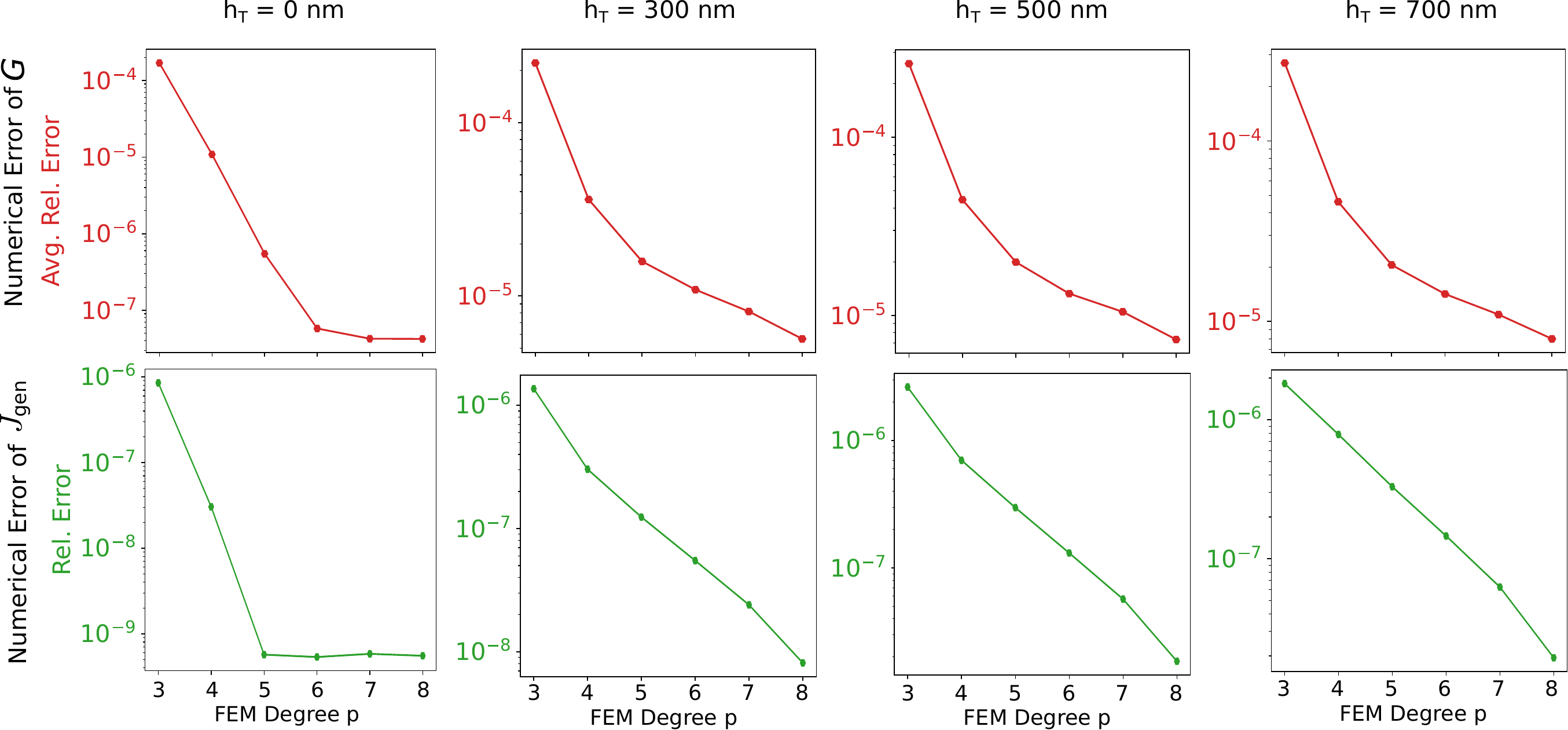}
  \caption{
      The numerical error of the finite element simulation with respect to the used polynomial degree $p$ and fixed maximum side length $h=0.5$. (a) The average relative error of the photogeneration rate $G$. (b) The relative error of the generated current density $J_{\text{gen}}$.
    }
  \label{supp:figs14}
\end{figure*}

The accuracy of the FEM simulations is primarily determined by two parameters: the polynomial degree $p$ and the maximum element side length $h$.
Higher polynomial degrees $p$ improve the approximation quality on each finite element, which, in turn, reduces the numerical error.
Similarly, decreasing the maximum element side length $h$ leads to a finer mesh and allows for a more accurate representation of the geometry and solution.
To quantify the error, we compute the relative error
$\epsilon_{rel} \coloneqq (u_k-u)/u$, where  $k\in{p,h}$.
Here, $u_k$ denotes the numerical solution computed with a given polynomial degree $p$ or mesh size $h$, and $u$ is a reference solution obtained with high $p$ or small $h$, respectively.
Figure \ref{supp:figs14} shows the average relative error in the photogeneration rate $G(\mathbf{x})$ as well as the relative error of the generated current density $J_{\text{gen}}$ in dependence of the polynomial degree $p$.


\newpage
\bibliographystyle{apsrev4-2}
\bibliography{lit}

\makeatletter\@input{xxMain.tex}\makeatother